\documentstyle[10pt,epsf,epsfig,dp_delphititle,oldlfont]{dp_delphi}
\makeindex
\def\DpPaperGroup{EP}
\def\DpPaperRef{2001-059}
\def\DpDate{1 August 2001}
\def\DpAuthors{DELPHI Collaboration}
\def\DpSubmit{(Accepted by Eur.Phys.J.C)}
\def\DpTitle{{\boldmath
        Search for Technicolor with DELPHI }}
\def\DpComment{ }
\def\DpEMail{ }

\newcommand{\qqg} {\mbox{$ {\mathrm q}\bar{\mathrm q}(\gamma) $}}

\newcommand{\rs}{\mbox{$\sqrt{s}$}}
\newcommand{\GeV} {\mbox{${\mathrm{GeV}} $}}

\newcommand{\Zz}{\mbox{${\rm Z}^0$}}

\begin{document}
\makeatletter
\newcount\@tempcntc
\def\@citex[#1]#2{\if@filesw\immediate\write\@auxout{\string\citation{#2}}\fi
  \@tempcnta\z@\@tempcntb\m@ne\def\@citea{}\@cite{\@for\@citeb:=#2\do
    {\@ifundefined
       {b@\@citeb}{\@citeo\@tempcntb\m@ne\@citea\def\@citea{,}{\bf ?}\@warning
       {Citation `\@citeb' on page \thepage \space undefined}}%
    {\setbox\z@\hbox{\global\@tempcntc0\csname b@\@citeb\endcsname\relax}%
     \ifnum\@tempcntc=\z@ \@citeo\@tempcntb\m@ne
       \@citea\def\@citea{,}\hbox{\csname b@\@citeb\endcsname}%
     \else
      \advance\@tempcntb\@ne
      \ifnum\@tempcntb=\@tempcntc
      \else\advance\@tempcntb\m@ne\@citeo
      \@tempcnta\@tempcntc\@tempcntb\@tempcntc\fi\fi}}\@citeo}{#1}}
\def\@citeo{\ifnum\@tempcnta>\@tempcntb\else\@citea\def\@citea{,}%
  \ifnum\@tempcnta=\@tempcntb\the\@tempcnta\else
   {\advance\@tempcnta\@ne\ifnum\@tempcnta=\@tempcntb \else \def\@citea{--}\fi
    \advance\@tempcnta\m@ne\the\@tempcnta\@citea\the\@tempcntb}\fi\fi}
 
\makeatother
\begin{titlepage}
\pagenumbering{roman}
\CERNpreprint{\DpPaperGroup}{\DpPaperRef} 
\date{{\small\DpDate}} 
\title{\DpTitle} 
\address{\DpAuthors} 
\begin{shortabs} 
\noindent
 Technicolor represents a viable alternative to the Higgs
mechanism for generating gauge boson masses. Searches for technicolor 
particles $\rho_T$ and $\pi_T$ have been performed in the data collected 
by the DELPHI experiment at LEP at centre-of-mass energies between 192 
and 208 GeV corresponding to an integrated luminosity of 452 pb$^{-1}$. Good 
agreement is observed with the SM expectation in all channels studied. 
This is translated into an excluded region in the $(M_{\pi_T},M_{\rho_T})$ 
plane. The $\rho_T$ production is excluded for all 
$90 < M_{\rho_T}<206.7$ GeV/c$^2$. Assuming a point-like interaction of 
the $\pi_T$ with gauge bosons, an absolute lower limit on the charged $\pi_T$ 
mass at 95\% CL is set at 79.8 GeV/c$^2$, independently of other parameters 
of the technicolor model.
\end{shortabs}
\vfill
\begin{center}
\DpSubmit \ \\ 
\DpComment \ \\
\DpEMail \ \\
\end{center}
\vfill
\clearpage
\headsep 10.0pt
\addtolength{\textheight}{10mm}
\addtolength{\footskip}{-5mm}
\begingroup
%
\newcommand{\DpName}[2]{\hbox{#1$^{\ref{#2}}$},\hfill}
\newcommand{\DpNameTwo}[3]{\hbox{#1$^{\ref{#2},\ref{#3}}$},\hfill}
\newcommand{\DpNameThree}[4]{\hbox{#1$^{\ref{#2},\ref{#3},\ref{#4}}$},\hfill}
\newskip\Bigfill \Bigfill = 0pt plus 1000fill
\newcommand{\DpNameLast}[2]{\hbox{#1$^{\ref{#2}}$}\hspace{\Bigfill}}
%
\footnotesize
\noindent
\DpName{J.Abdallah}{LPNHE}
\DpName{P.Abreu}{LIP}
\DpName{W.Adam}{VIENNA}
\DpName{P.Adzic}{DEMOKRITOS}
\DpName{T.Albrecht}{KARLSRUHE}
\DpName{T.Alderweireld}{AIM}
\DpName{R.Alemany-Fernandez}{CERN}
\DpName{P.P.Allport}{LIVERPOOL}
\DpName{S.Almehed}{LUND}
\DpName{T.Almendinger}{KARLSRUHE}
% ...
\normalsize
\endgroup
\titlefoot{Department of Physics and Astronomy, Iowa State
     University, Ames IA 50011-3160, USA
    \label{AMES}}
\titlefoot{Physics Department, Universiteit Antwerpen,
     Universiteitsplein 1, B-2610 Antwerpen, Belgium \\
     \indent~~and IIHE, ULB-VUB,
     Pleinlaan 2, B-1050 Brussels, Belgium \\
     \indent~~and Facult\'e des Sciences,
     Univ. de l'Etat Mons, Av. Maistriau 19, B-7000 Mons, Belgium
    \label{AIM}}
\titlefoot{Physics Laboratory, University of Athens, Solonos Str.
     104, GR-10680 Athens, Greece
    \label{ATHENS}}
\titlefoot{Department of Physics, University of Bergen,
     All\'egaten 55, NO-5007 Bergen, Norway
    \label{BERGEN}}
\titlefoot{Dipartimento di Fisica, Universit\`a di Bologna and INFN,
     Via Irnerio 46, IT-40126 Bologna, Italy
    \label{BOLOGNA}}
\titlefoot{Centro Brasileiro de Pesquisas F\'{\i}sicas, rua Xavier Sigaud 150,
     BR-22290 Rio de Janeiro, Brazil \\
     \indent~~and Depto. de F\'{\i}sica, Pont. Univ. Cat\'olica,
     C.P. 38071 BR-22453 Rio de Janeiro, Brazil \\
     \indent~~and Inst. de F\'{\i}sica, Univ. Estadual do Rio de Janeiro,
     rua S\~{a}o Francisco Xavier 524, Rio de Janeiro, Brazil
    \label{BRASIL}}
\titlefoot{Coll\`ege de France, Lab. de Physique Corpusculaire, IN2P3-CNRS,
     FR-75231 Paris Cedex 05, France
    \label{CDF}}
\titlefoot{CERN, CH-1211 Geneva 23, Switzerland
    \label{CERN}}
\titlefoot{Institut de Recherches Subatomiques, IN2P3 - CNRS/ULP - BP20,
     FR-67037 Strasbourg Cedex, France
    \label{CRN}}
\titlefoot{Now at DESY-Zeuthen, Platanenallee 6, D-15735 Zeuthen, Germany
    \label{DESY}}
\titlefoot{Institute of Nuclear Physics, N.C.S.R. Demokritos,
     P.O. Box 60228, GR-15310 Athens, Greece
    \label{DEMOKRITOS}}
\titlefoot{Dipartimento di Fisica, Universit\`a di Genova and INFN,
     Via Dodecaneso 33, IT-16146 Genova, Italy
    \label{GENOVA}}
\titlefoot{Institut des Sciences Nucl\'eaires, IN2P3-CNRS, Universit\'e
     de Grenoble 1, FR-38026 Grenoble Cedex, France
    \label{GRENOBLE}}
\titlefoot{Helsinki Institute of Physics, HIP,
     P.O. Box 9, FI-00014 Helsinki, Finland
    \label{HELSINKI}}
\titlefoot{Joint Institute for Nuclear Research, Dubna, Head Post
     Office, P.O. Box 79, RU-101 000 Moscow, Russian Federation
    \label{JINR}}
\titlefoot{Institut f\"ur Experimentelle Kernphysik,
     Universit\"at Karlsruhe, Postfach 6980, DE-76128 Karlsruhe,
     Germany
    \label{KARLSRUHE}}
\titlefoot{Institute of Nuclear Physics,Ul. Kawiory 26a,
     PL-30055 Krakow, Poland
    \label{KRAKOW1}}
\titlefoot{Faculty of Physics and Nuclear Techniques, University of Mining
     and Metallurgy, PL-30055 Krakow, Poland
    \label{KRAKOW2}}
\titlefoot{Universit\'e de Paris-Sud, Lab. de l'Acc\'el\'erateur
     Lin\'eaire, IN2P3-CNRS, B\^{a}t. 200, FR-91405 Orsay Cedex, France
    \label{LAL}}
\titlefoot{School of Physics and Chemistry, University of Landcaster,
     Lancaster LA1 4YB, UK
    \label{LANCASTER}}
\titlefoot{LIP, IST, FCUL - Av. Elias Garcia, 14-$1^{o}$,
     PT-1000 Lisboa Codex, Portugal
    \label{LIP}}
\titlefoot{Department of Physics, University of Liverpool, P.O.
     Box 147, Liverpool L69 3BX, UK
    \label{LIVERPOOL}}
\titlefoot{LPNHE, IN2P3-CNRS, Univ.~Paris VI et VII, Tour 33 (RdC),
     4 place Jussieu, FR-75252 Paris Cedex 05, France
    \label{LPNHE}}
\titlefoot{Department of Physics, University of Lund,
     S\"olvegatan 14, SE-223 63 Lund, Sweden
    \label{LUND}}
\titlefoot{Universit\'e Claude Bernard de Lyon, IPNL, IN2P3-CNRS,
     FR-69622 Villeurbanne Cedex, France
    \label{LYON}}
\titlefoot{Dipartimento di Fisica, Universit\`a di Milano and INFN-MILANO,
     Via Celoria 16, IT-20133 Milan, Italy
    \label{MILANO}}
\titlefoot{Dipartimento di Fisica, Univ. di Milano-Bicocca and
     INFN-MILANO, Piazza delle Scienze 2, IT-20126 Milan, Italy
    \label{MILANO2}}
\titlefoot{IPNP of MFF, Charles Univ., Areal MFF,
     V Holesovickach 2, CZ-180 00, Praha 8, Czech Republic
    \label{NC}}
\titlefoot{NIKHEF, Postbus 41882, NL-1009 DB
     Amsterdam, The Netherlands
    \label{NIKHEF}}
\titlefoot{National Technical University, Physics Department,
     Zografou Campus, GR-15773 Athens, Greece
    \label{NTU-ATHENS}}
\titlefoot{Physics Department, University of Oslo, Blindern,
     NO-1000 Oslo 3, Norway
    \label{OSLO}}
\titlefoot{Dpto. Fisica, Univ. Oviedo, Avda. Calvo Sotelo
     s/n, ES-33007 Oviedo, Spain
    \label{OVIEDO}}
\titlefoot{Department of Physics, University of Oxford,
     Keble Road, Oxford OX1 3RH, UK
    \label{OXFORD}}
\titlefoot{Dipartimento di Fisica, Universit\`a di Padova and
     INFN, Via Marzolo 8, IT-35131 Padua, Italy
    \label{PADOVA}}
\titlefoot{Rutherford Appleton Laboratory, Chilton, Didcot
     OX11 OQX, UK
    \label{RAL}}
\titlefoot{Dipartimento di Fisica, Universit\`a di Roma II and
     INFN, Tor Vergata, IT-00173 Rome, Italy
    \label{ROMA2}}
\titlefoot{Dipartimento di Fisica, Universit\`a di Roma III and
     INFN, Via della Vasca Navale 84, IT-00146 Rome, Italy
    \label{ROMA3}}
\titlefoot{DAPNIA/Service de Physique des Particules,
     CEA-Saclay, FR-91191 Gif-sur-Yvette Cedex, France
    \label{SACLAY}}
\titlefoot{Instituto de Fisica de Cantabria (CSIC-UC), Avda.
     los Castros s/n, ES-39006 Santander, Spain
    \label{SANTANDER}}
\titlefoot{Inst. for High Energy Physics, Serpukov
     P.O. Box 35, Protvino, (Moscow Region), Russian Federation
    \label{SERPUKHOV}}
\titlefoot{J. Stefan Institute, Jamova 39, SI-1000 Ljubljana, Slovenia
     and Laboratory for Astroparticle Physics,\\
     \indent~~Nova Gorica Polytechnic, Kostanjeviska 16a, SI-5000 Nova Gorica, Slovenia, \\
     \indent~~and Department of Physics, University of Ljubljana,
     SI-1000 Ljubljana, Slovenia
    \label{SLOVENIJA}}
\titlefoot{Fysikum, Stockholm University,
     Box 6730, SE-113 85 Stockholm, Sweden
    \label{STOCKHOLM}}
\titlefoot{Dipartimento di Fisica Sperimentale, Universit\`a di
     Torino and INFN, Via P. Giuria 1, IT-10125 Turin, Italy
    \label{TORINO}}
\titlefoot{Dipartimento di Fisica, Universit\`a di Trieste and
     INFN, Via A. Valerio 2, IT-34127 Trieste, Italy \\
     \indent~~and Istituto di Fisica, Universit\`a di Udine,
     IT-33100 Udine, Italy
    \label{TU}}
\titlefoot{Univ. Federal do Rio de Janeiro, C.P. 68528
     Cidade Univ., Ilha do Fund\~ao
     BR-21945-970 Rio de Janeiro, Brazil
    \label{UFRJ}}
\titlefoot{Department of Radiation Sciences, University of
     Uppsala, P.O. Box 535, SE-751 21 Uppsala, Sweden
    \label{UPPSALA}}
\titlefoot{IFIC, Valencia-CSIC, and D.F.A.M.N., U. de Valencia,
     Avda. Dr. Moliner 50, ES-46100 Burjassot (Valencia), Spain
    \label{VALENCIA}}
\titlefoot{Institut f\"ur Hochenergiephysik, \"Osterr. Akad.
     d. Wissensch., Nikolsdorfergasse 18, AT-1050 Vienna, Austria
    \label{VIENNA}}
\titlefoot{Inst. Nuclear Studies and University of Warsaw, Ul.
     Hoza 69, PL-00681 Warsaw, Poland
    \label{WARSZAWA}}
\titlefoot{Fachbereich Physik, University of Wuppertal, Postfach
     100 127, DE-42097 Wuppertal, Germany
    \label{WUPPERTAL}}
\addtolength{\textheight}{-10mm}
\addtolength{\footskip}{5mm}
\clearpage
\headsep 30.0pt
\end{titlepage}
%
\pagenumbering{arabic} 
\setcounter{footnote}{0} %
\large
\section[]{Introduction}

In spite of outstanding theoretical and experimental achievements, 
particle physicists have not been able to decide which mechanism 
creates mass.
It is a common belief that such a
mechanism will be characterised by the observation of at least a scalar
particle. Whether this object is elementary (as in the SM or MSSM scenario), 
composite (as in the technicolor scenario), or too heavy to be observed 
as a particle remains uncertain. 

This paper presents a systematic search for the particles predicted by the
technicolor model. Section 3 briefly recalls the framework of the
technicolor (TC) model and reviews the possible signals 
which can be observed at LEP2.
Section 4 describes the direct search for technipions performed with the 
DELPHI detector using the data collected in 1999 and 2000. 
Section 5 presents complementary searches for technirho ($\rho_T$) production
for $M_{\rho_T} < \sqrt{s}$ in the region of higher technipion masses.
Section 6 summarises the combined results.

\section[]{Data Sample}

The detailed description of the DELPHI detector can be found 
elsewhere~\cite{delphi}.
For the search for $\pi_T$ production, the statistics of DELPHI taken in 1999
for $\sqrt{s}$ between 192 and 202 GeV and in 2000 for 
$\sqrt{s}$ between 202 and 208 GeV are used. The integrated luminosity
is about 228 pb$^{-1}$ for data taken in 1999 and 224 pb$^{-1}$ for data
taken in 2000. 
In addition, the available DELPHI $e^+e^- \to W^+W^-$~\cite{ww-cs} and 
$e^+e^- \to q \bar{q} (\gamma)$~\cite{qq-cs}  cross-section measurements 
are used to estimate a possible contribution from technicolor production.

Simulated events are produced with the DELPHI simulation program
DELSIM and are passed through the same reconstruction chain as
the data. To simulate the Standard Model (SM) backgrounds,
the generator EXCALIBUR~\cite{exc} is used for 4-fermion final states, 
PYTHIA~\cite{pyth} for the process $e^+e^- \rightarrow q \bar{q} (+n \gamma)$,
and TWOGAM~\cite{gg} for two-photon interactions.
The technicolor production signal is simulated 
using a special generator~\cite{mrenna} included in the PYTHIA package.

\section[]{The Technicolor scheme at LEP}
\vskip 0.5 cm
The technicolor model provides an elegant scheme to generate W/Z masses. 
These bosons are seen as condensates of a new family of quarks (the 
techniquarks) which obey a QCD-like interaction with an effective scale 
$\Lambda_{TC}$ much larger than $\Lambda_{QCD}$. 
It also predicts heavy ($>$~1 TeV)
vector mesons which cannot be observed at LEP2. 

It is well known, however, that this scheme encounters several problems. 
It cannot correctly generate fermion masses and, in its simplest version, 
it contradicts the LEP1 precision measurements since it gives positive 
contributions to the S parameter. In technicolor models with QCD-like
dynamics, $S \sim 0.45$ is expected for an isodoublet of technifermions,
while the precise measurements give: $S = -0.07 \pm 0.11$~\cite{ref1}.

Extensions~\cite{ref2} have been worked out which solve these  
problems at the price of losing predictive power. These schemes depart from
the straightforward analogy with QCD, with the usual asymptotic 
freedom behaviour.
It turns out that perturbative calculations do not work 
(``walking technicolor''), and therefore
the theory cannot be fully tested by precision measurements. 

These extensions call for a large number N$_D$ of technidoublets~\cite{ref3},
and therefore for additional scalar ($\pi_T$, $\pi'_T$) and
vector ($\rho_T$, $\omega_T$) mesons. 
These can be light enough to be observed at LEP2 or the Tevatron. 
Our searches for technicolor production assume the
theoretical model given in~\cite{lane}. 

The main $\rho_T$ decay modes are $\rho_T \to \pi_T \pi_T$,
$W_L \pi_T$, $W_L W_L$, $f_i \bar{f_i}$ and $\pi_T^0 \gamma$,
where $W_L$ is the longitudinal component of the $W$ boson.
For $ M_{\rho_T} > 2 M_{\pi_T}$ the decay $\rho_T \to \pi_T \pi_T$
is dominant, while for $M_{\rho_T} < 2 M_{\pi_T}$ the decay rates 
depend on many model parameters. In all cases the total $\rho_T$ width 
for $M_{\rho_T} < 200$ GeV/c$^2$ is predicted to be of the order of 10 GeV 
if any of the channels $\rho_T \to \pi_T \pi_T$, $\pi_T W_L$, 
$W_L W_L$ is open, and below 1 GeV if all of them are closed. 
For $\omega_T$ the main decay modes are $\omega_T \to \pi_T 
\pi_T \pi_T$, $\pi_T \pi_T W_L$, etc. If these
decay modes are forbidden kinematically, then its dominant decay is 
$\omega_T \to \pi^0_T \gamma$. 
By analogy with QCD it is supposed that 
$M_{\rho_T} \simeq M_{\omega_T}$
and 
$M_{\pi^0_T} \simeq M_{\pi^\pm_T}$.


Following~\cite{lane}, technipions are assumed to decay as
$\pi_T^+ \to c \bar{b}$, $c \bar{s}$ and $\tau^+ \nu_{\tau}$;
and $\pi_T^0 \to b \bar{b}$, $c \bar{c}$ and $\tau^+ \tau^-$.
The width $\Gamma(\pi_T \to \bar{f'} f)$ is proportional to
$(m_f +m_{f'})^2$, therefore the $b$-quark is produced in $\sim90$\% 
of $\pi_T$ decays. The total $\pi_T$ width is less than 1 GeV.
These properties are extensively used in the following.

The $\rho_T$ coupling to the photon and $Z^0$ is proportional to $Q_U-Q_D$, 
where $Q_U$ and $Q_D$ are the charges of $U$ and $D$ techniquarks.
The value $Q_U-Q_D$ has to be one to avoid triangle anomalies.
Therefore, for $M_{\rho_T} < \sqrt{s}$, it can be produced on mass shell 
in $e^+e^-$ interactions through the radiative return process
and its production cross-section is 
independent of the values chosen for $Q_U$ and $Q_D$.
It can then be observed as a narrow resonance in the corresponding mass 
distribution. 
The radiative return production rate normalised to the point-like 
cross-section is given approximately by:
\begin{equation}
 R(e^+e^- \to \rho_T (\gamma)) \simeq \ln(s/m^2_e)
\frac{\Gamma^{e^+e^-}_{\rho_T}/M_{\rho_T}}{\Gamma^{e^+e^-}_Z/M_Z} 
\frac{1}{1-M^2_{\rho_T}/s} 
\label{eq1}
\end{equation}

\vspace*{0.2cm}

In addition, $\omega_T$ can also couple to $e^+e^-$ 
provided $Q_U+Q_D$ is non-zero.
The following always supposes that the final state $\pi_T^0 \gamma$ 
can be produced through both $\rho_T$ and $\omega_T$.

Technipions can also be produced at LEP through virtual $\rho_T$
exchange. The analyses presented below use the off-shell processes
$e^+e^- \to \rho_T^* \to (\pi^+_T \pi^-_T$, $\pi^+_T W^-_L)$ and 
$e^+e^- \to (\rho_T^*, \omega_T^*) \to \pi^0_T \gamma$ 
to search for virtual $\rho_T$ production if $M_{\rho_T}>\sqrt{s}$.
The cross-sections of these processes normalised  to the
point-like cross-section, derived for $e^+e^-$ interactions 
from equations given in~\cite{lane}, are:
\begin{eqnarray}
\label{eq2}
  R(e^+e^- \to \rho_T^* \rightarrow a^+b^-) & = &
\frac{[|A_{eL}(s)|^2+|A_{eR}(s)|^2]~\lambda(M_a,M_b)^{3/2}~C_{ab}}
{8(1-s/M^2_{\rho_T})^2}; \\
\label{eq3}
  R(e^+e^- \to (\rho_T^*, \omega_T^*) \rightarrow \pi_T \gamma) & = &
\frac{[|C_{eL}(s)|^2+|C_{eR}(s)|^2]~\lambda(M_{\pi_T}, 0)^{3/2}~\cos^2\chi}
{16(1-s/M^2_{\rho_T})^2} \times \nonumber \\
&  & \frac{\alpha \cdot (Q_U+Q_D)^2 \cdot s} 
{\alpha_{\rho_T} \cdot M_V^2} 
\end{eqnarray}
In these equations $a,b = \pi_T, W_L$;~~
$C_{ab} = \cos^4 \chi$ for $\pi^+_T \pi^-_T$, 
$2\cos^2\chi \sin^2\chi$ for $\pi^+_T W^-_L$,
and $\sin^4 \chi$ for $W^+_L W^-_L$;~~
and the angle $\chi$ reflects the mixing between $\pi_T$ and $W_L$ with
\begin{equation}
\sin^2\chi = 1/N_D
\label{eq:chi}
\end{equation}
The values $A_{eL,R}$ and $C_{eL,R}$ in (\ref{eq2}) and (\ref{eq3})
are given by:
\begin{eqnarray}
A_{eL,R}(s) & = & Q_e+\frac{2\cos2\theta_W}{\sin^22\theta_W}
(T_{3eL,R}-Q_e\sin^2\theta_W)BW_Z, \\
C_{eL,R}(s) & = & 2Q_e-\frac{2}{\sin^22\theta_W}
(T_{3eL,R}-Q_e\sin^2\theta_W)BW_Z, \\
 BW_Z & = & \frac{s}{s-M^2_Z+i\sqrt{s}\Gamma_Z},
\end{eqnarray}
where $Q_e = -1$, $T_{3eL} = -1/2$, $T_{3eR} = 0$.
The phase space suppression factor $\lambda(M_a,M_b)$ is:
\begin{equation}
\label{eq7}
 \lambda(M_a,M_b) =(1-M^2_a/s-M^2_b/s)^2-4M^2_aM^2_b/s^2.
\end{equation}

Note that for a highly virtual $\rho_T$ contribution, even for 
$M^2_{\rho_T}\rightarrow\infty $, the value of 
$R(e^+e^- \to \rho_T\rightarrow a^+b^-)$ remains finite. If 
the $Z$ contributions are ignored, expressions (\ref{eq2}-\ref{eq7}) lead to 
$R(e^+e^- \to \rho_T\to a^+b^-)\sim \lambda(M_a,M_b)^{3/2} C_{ab}/4$, 
as expected for a point-like coupling of a photon to $\pi^+_T\pi^-_T$. 
This correct behaviour results from our choice of the $\rho_T$ propagator. 
This feature is important,  as it allows LEP to be sensitive to 
a light $\pi_T$ even if the $\rho_T$ is very heavy.

\begin{figure}[tbh]
  \begin{center}
    \psfig{figure=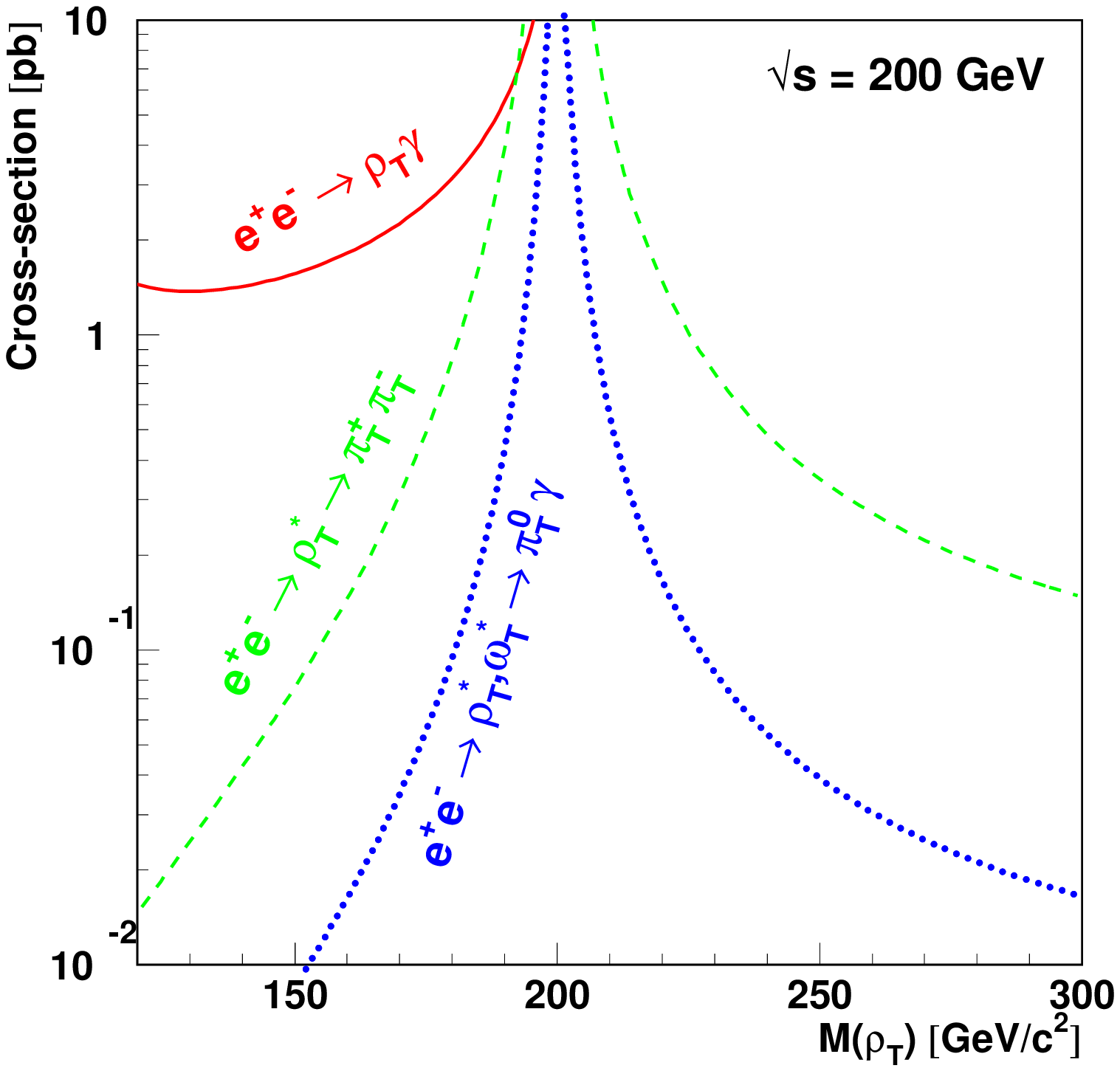,width=17cm}
    \caption[]{Technicolor production cross-sections at LEP for some
      typical parameter values:
      $M_{\pi_T} = 90$ GeV/c$^2$, $M_V = 200$ GeV/c$^2$, $N_D = 9$, 
      $(Q_U+Q_D) = 4/3$, and $\alpha_{\rho_T} = 2.91(3/N_{TC})$ with 
      $N_{TC} = 4$.
      }
    \label{fig:tccs}
  \end{center}
\end{figure}

The processes $e^+e^- \to \rho_T^* \to (\pi^+_T \pi^-_T$, $\pi^+_T W^-_L)$
depend on 3 quantities, namely $M_{\pi_T}$, $M_{\rho_T}$ and $N_D$.
Three additional parameters, namely the technicolor coupling
constant $\alpha_{\rho_T}$, the sum of charges of the
technicolor doublet $Q_U + Q_D$, and the mass scale $M_V$ are introduced 
to describe $e^+e^- \to (\rho_T^*, \omega_T^*) \to \pi^0_T \gamma$.
Figure \ref{fig:tccs} shows the cross-sections of processes 
(\ref{eq1}-\ref{eq3}) for some typical
parameter values  proposed in~\cite{lane}:
$M_{\pi_T} = 90$ GeV/c$^2$, $M_V = 200$ GeV/c$^2$, $N_D = 9$,
$(Q_U+Q_D) = 4/3$. It is assumed that the symmetry group, under
which the technifermions transform as fundamental, is $SU(N_{TC})$
with $N_{TC} = 4$ and that $\alpha_{\rho_T} = 2.91(3/N_{TC})$.

It can be seen that the production cross-section of technicolor
objects is expected to be reasonably high for a wide range of $M_{\rho_T}$
values, making the search at LEP possible, but that the process (\ref{eq3}), 
giving the $\pi^0_T \gamma$ final state, depends strongly on the three 
additional parameters, and can even become zero for $(Q_U + Q_D) = 0$.

This paper reports searches for $\rho_T$ with $M_{\rho_T} < \sqrt{s}$ 
in all decay modes in process (\ref{eq1}), 
for $\pi_T^+ \pi_T^-$ and $\pi_T^+ W^-_L$ final states in 
process (\ref{eq2}),
and for $\pi_T\gamma$ in process (\ref{eq3}).
It is assumed that $M_{\rho_T}>90$ GeV/c$^2$ and $M_{\pi_T} > 45$ GeV/c$^2$,
supposing that the $\rho_T$ and $\pi_T$ with smaller masses would be detected 
in precise measurements at LEP1.
The CDF experiment at the Tevatron~\cite{ref4} 
has already published results of a search for these particles. 

\section[]{Search for \boldmath $\pi_T$ in $e^+ e^- \to \rho_T^{*} \to 
(W_L \pi_T, \pi_T\pi_T)$}

\label{pipi}

If the ${\pi_T}$ is light enough, $W^+_L\pi^-_T$ or even $\pi^+_T\pi^-_T$ 
final states can be produced in process (\ref{eq2}).
These can provide striking signatures because
technipions are expected~\cite{lane}
to decay into the heaviest fermions.
Charged technipions therefore prefer final states  with a $b$ quark, 
which can be separated from the W bosons by applying b-tagging. 

\subsection[]{Search in the 4-jet Final State}

\label{sec41}

Events originating from the signal contain mainly one or two b-quarks
and one or two c-quarks, while the background from $W^+ W^-$ 
contains very few  b-quarks. This situation is similar to that 
in the Higgs search in 4 jet final states, therefore the same jet clustering
algorithm using the DURHAM method~\cite{durham} and the same b-tagging 
procedure~\cite{fastpap} are applied.
The analysis starts with 
the four-jet preselection described in~\cite{pap97},
which aims to eliminate the radiative and $\gamma \gamma$ events and to
reduce the QCD and $\Zz\gamma^*$ background.

The $q\bar{q}(\gamma)$ and 4-fermion backgrounds remaining after the 
preselection have to be reduced further. For this purpose different shape and 
${\mathrm b}$-tagging variables have been investigated, assuming that 
the analysis should be sensitive and  keep a reasonable efficiency for 
a wide range of the $\pi_T$ mass from $\sim45$ GeV/c$^2$ up to the 
kinematical limit. 

Finally, 12 variables are selected for this analysis and the
final discriminant variable is defined as the output of a
neural network (NN).
There are two ${\mathrm b}$-tagging variables intended 
to reduce the $W^+ W^-$ background: one of them ($x_b$) is computed as 
the sum of the two highest jet ${\mathrm b}$-tagging variables~\cite{btag}, 
and the other is the sum of the four jet ${\mathrm b}$-tagging variables.
Seven shape variables are used to reduce the \qqg\ contamination. 
They  are 
the sum of the second and fourth Fox-Wolfram moments,
the product of the minimum jet energy and the minimum opening angle between
any two jets,
the event thrust,
the sum of  the four lowest angles between any pair of jets in the event,
the minimal di-jet mass,
and the minimal $y_{cut}$ values
for which the event is clustered into 4 jets ($y_{34}$)
and into 5 jets ($y_{45}$). Finally, three more variables 
take into account the two-boson event topology. To define them the event 
is forced into four jets, a five constraint fit requiring conservation 
of energy and momentum and equal masses of opposite jet pairs is applied to 
all possible jet pairings, and the pairing giving the smallest 
value of the fit $\chi^2_{5C}$ is selected. 
The variables then included in the neural network are 
the smallest $\chi^2_{5C}$,
the production angle of the jet pair, 
and the angle between the planes defined by the two jet pairs. 

The resulting NN output provides good background suppression and high
selection efficiency over a wide range of $M_{\pi_T}$.
As an example, Table ~\ref{tabs200} gives the $\pi_T \pi_T$ and
$W_L \pi_T$ efficiencies for different ${\pi_T}$ masses
obtained when selecting events with NN output $>0.3$. 

\begin{figure}[htbp]
\begin{center}
\begin{tabular}{c}
\epsfig{figure=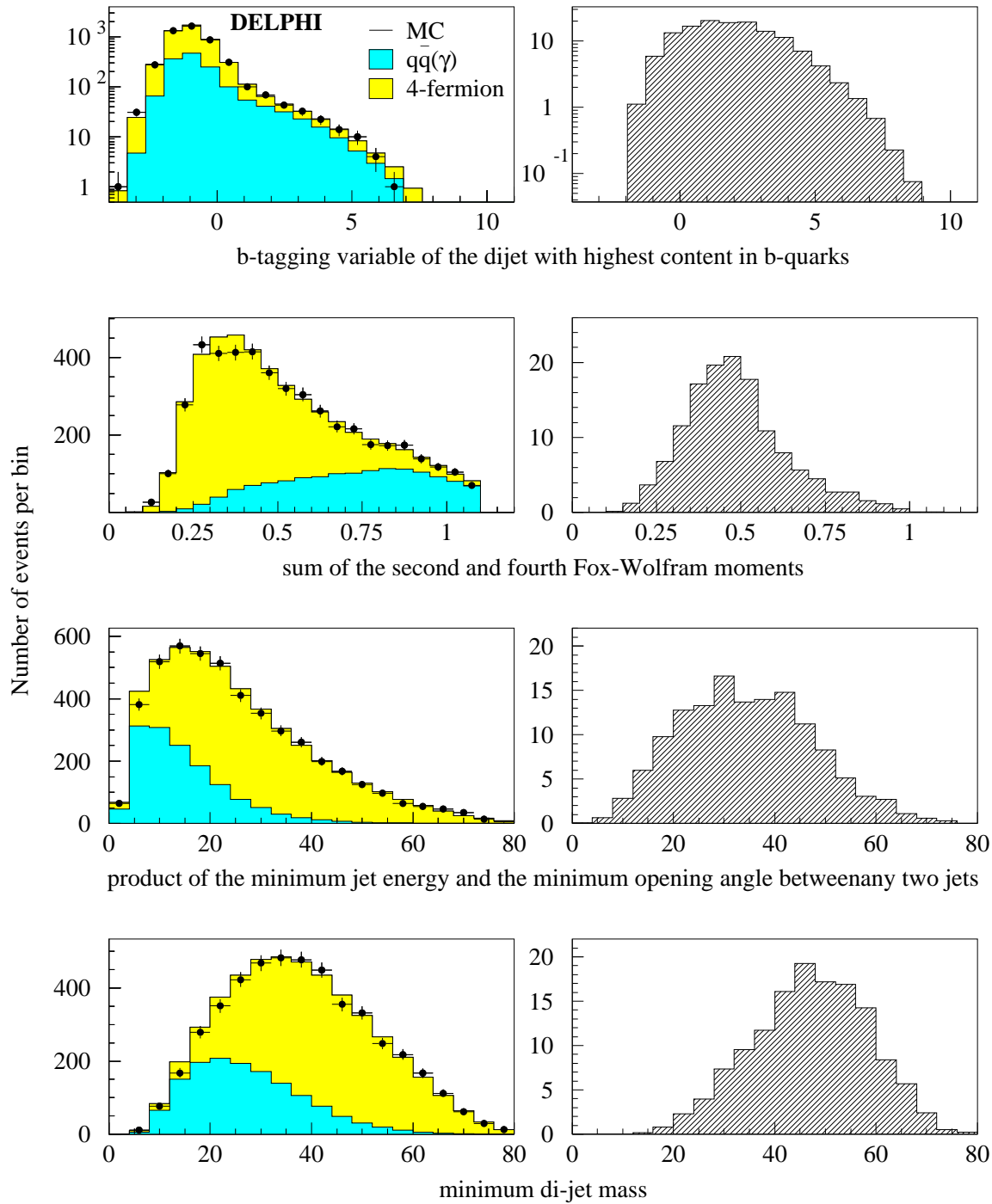,width=17.cm} 
\end{tabular}
\caption[] { Search in the 4-jet final state: 
distributions after preselection of the 
${\mathrm b}$-tagging variable, $H_2+H_4$, 
the product of the minimum jet energy
and the minimum opening angle between any two jets.
The plots on the left show the data (points)
and the expected SM backgrounds (histograms) for the full DELPHI statistics
at $\sqrt{s}=192-208$ GeV.
Those on the right show the 
technicolor signal expected in the channel $e^+e^-\to \pi_T\pi_T$ if 
$M_{\pi_T}$ = 99 GeV/c$^2$. The signal normalisation corresponds to 
$M_{\rho_T}=220$ GeV/c$^2$, $N_D=9$ and the integrated luminosity
collected at $\sqrt{s}$=192-208 GeV.}
\label{fig:pre4j}
\end{center}
\end{figure}

\begin{figure}[htbp]
\begin{center}
\begin{tabular}{c}
\epsfig{figure=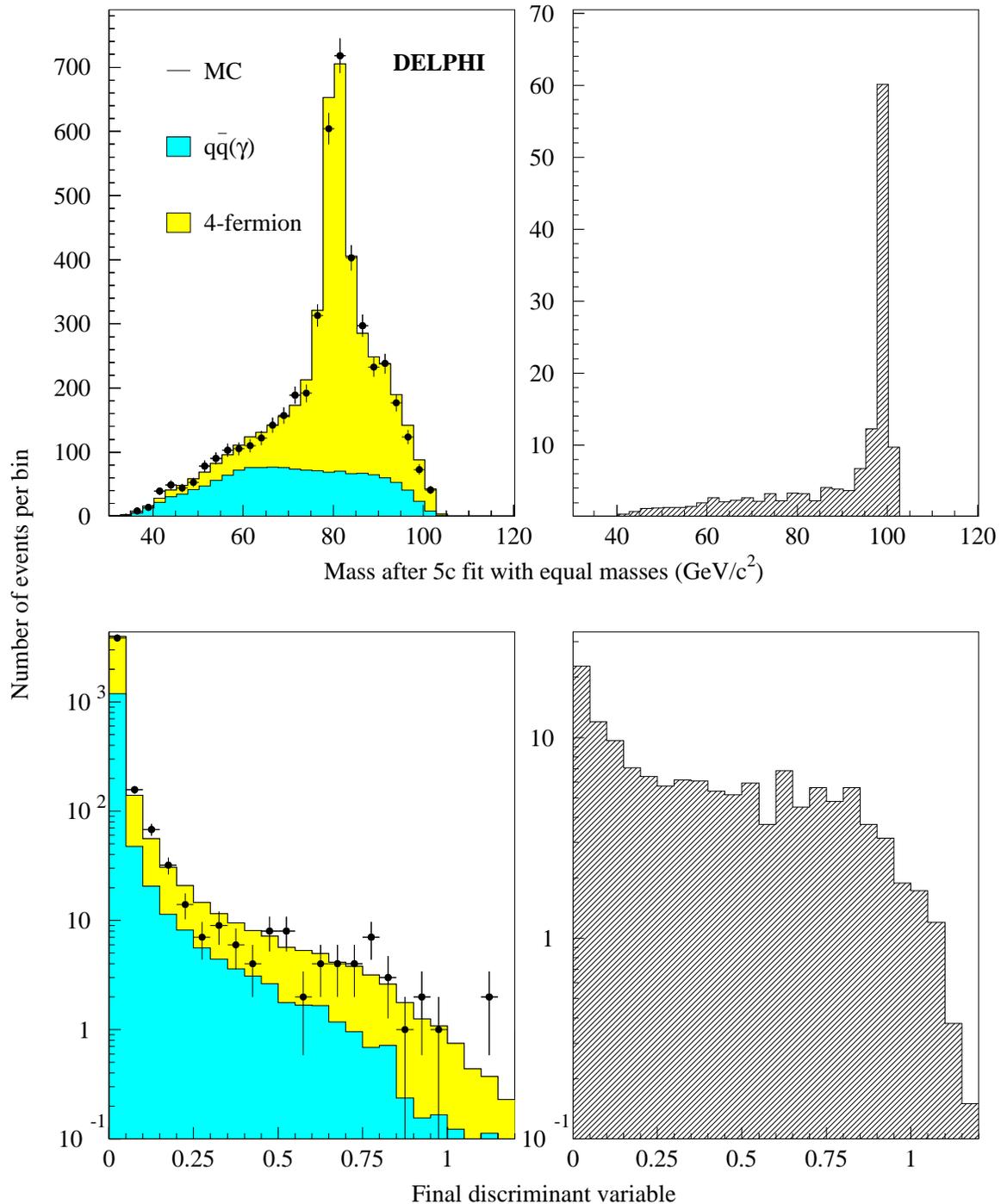,width=17.cm} 
\end{tabular}
\caption[] { Search in the 4-jet final state: distributions of the 
mass and final discriminant  variable after preselection. 
The plots on the left show the data (points)
and the expected SM backgrounds (histograms) for the full DELPHI statistics
at $\sqrt{s}=192-208$ GeV.
Those on the right show the 
technicolor signal in $e^+e^-\to \pi_T\pi_T$ expected if 
$M_{\pi_T}$ = 99 GeV/c$^2$. The signal normalisation corresponds to 
$M_{\rho_T}=220$ GeV/c$^2$, $N_D=9$ and the integrated luminosity
collected at $\sqrt{s}$=192-208 GeV.}
\label{fig:pre4j-mass}
\end{center}
\end{figure}

\begin{figure}[htbp]
\begin{center}
\epsfig{figure=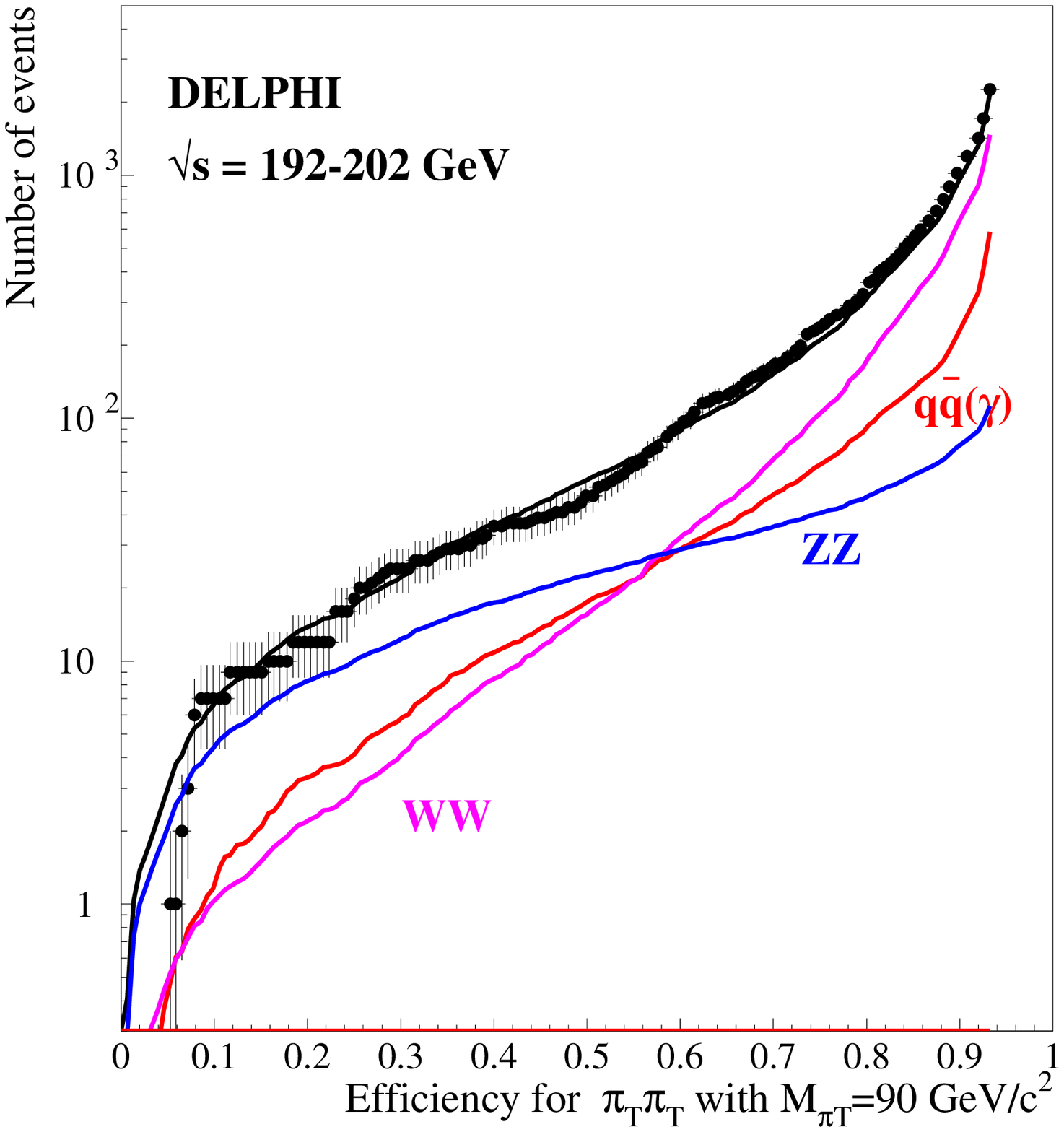,width=10.cm} 
\epsfig{figure=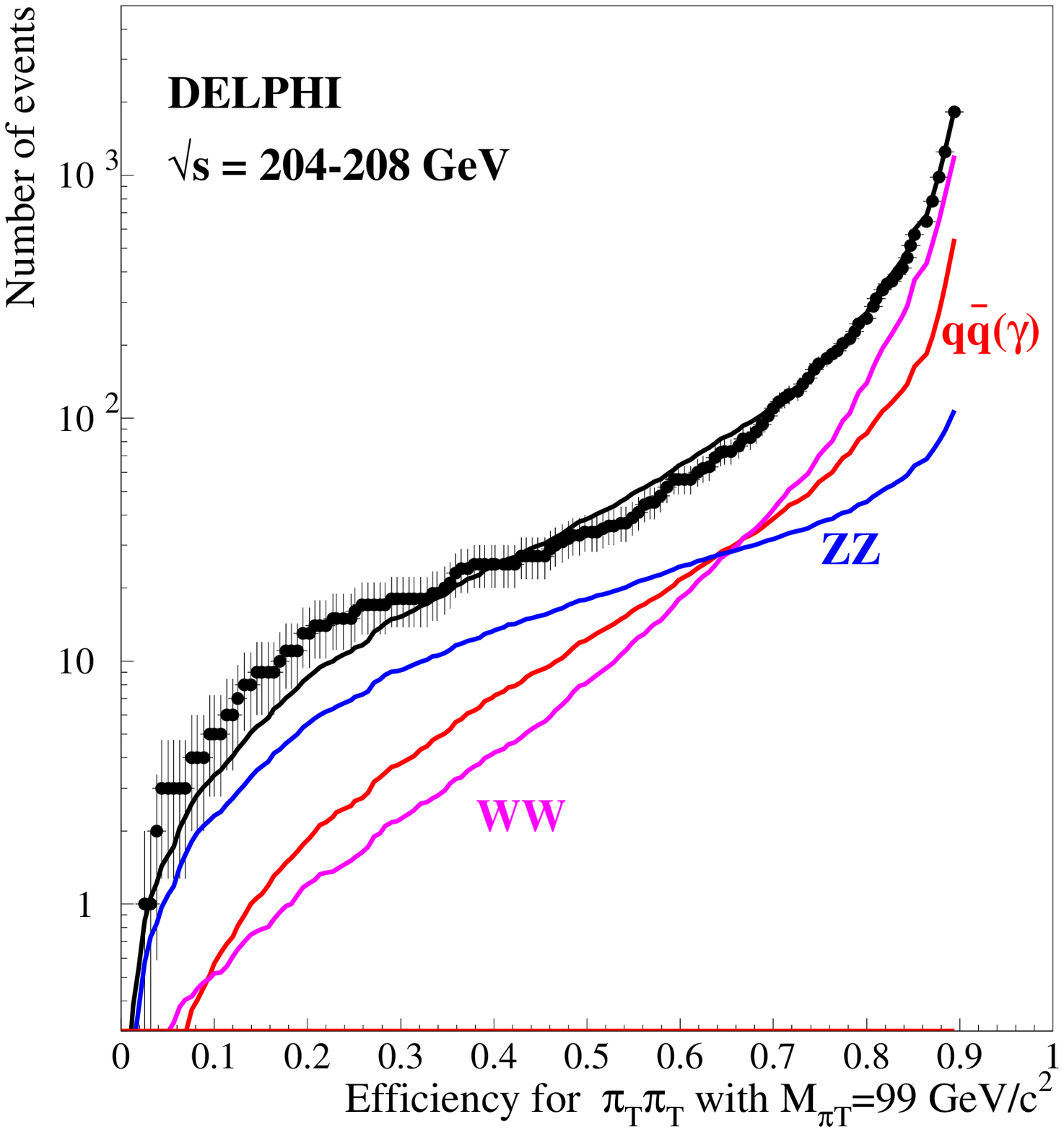,width=10.cm} 
\caption[] { Search in the 4-jet final state: numbers of data events (points) 
  and expected SM background events (curves) as a
  function of the $\pi_T\pi_T$ signal efficiency,
  varied by varying the cut on the neural network variable.
  The different background contributions are shown 
  both separately and combined.
  The two plots show the two different years of data taking considered.}
\label{fig:eff}
\end{center}
\end{figure}

\begin{figure}[htbp]
\begin{center}
\begin{tabular}{c}
\epsfig{figure=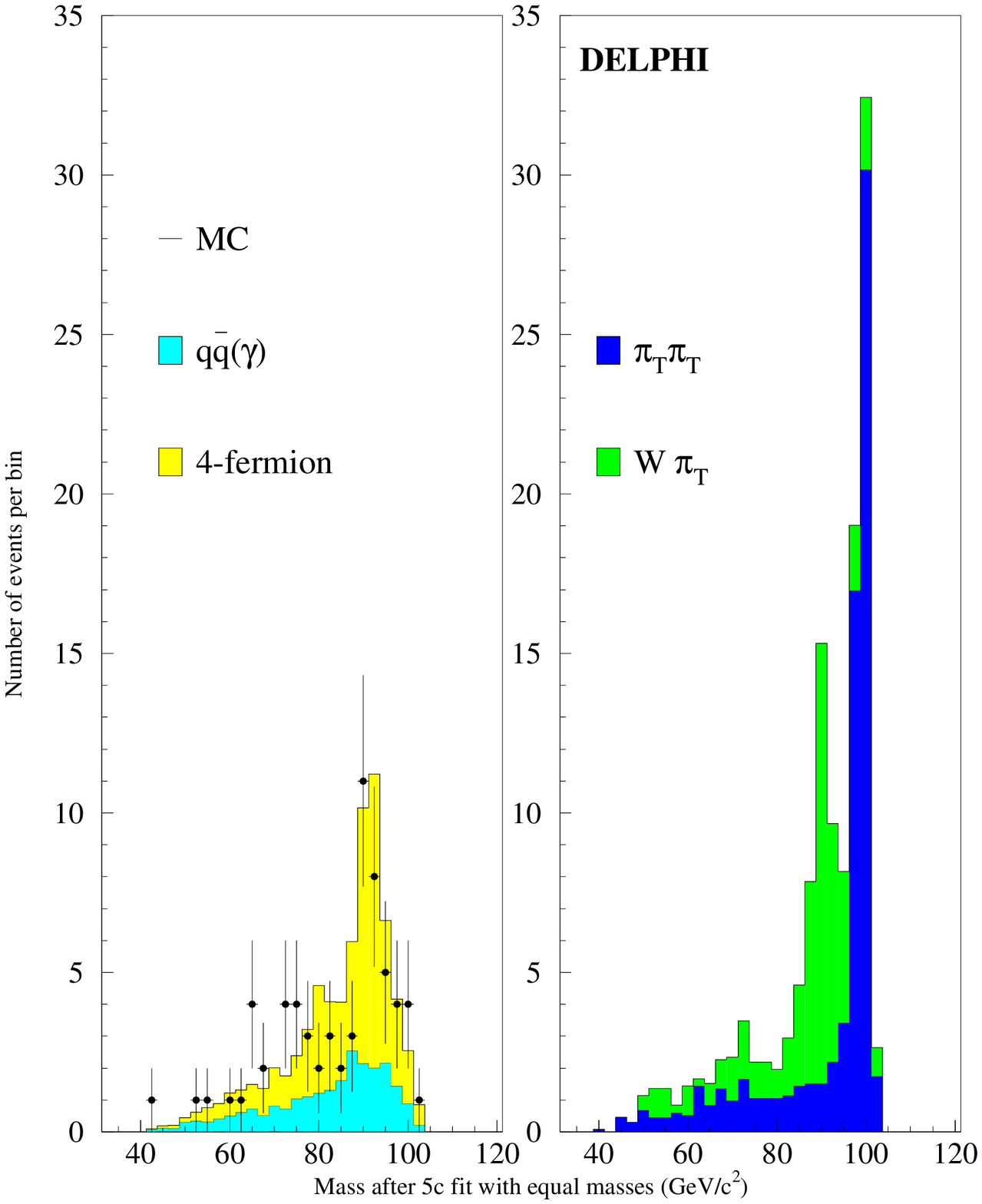,width=17.cm} 
\end{tabular}
\caption[] { Search in the 4-jet final state: $M_{5C}$ mass distributions for 
the NN analysis with the cut on NN output $>0.30$. The plot on the
left shows the data (points) and the expected SM backgrounds (histograms)
for the full DELPHI statistics at $\sqrt{s}=192-208$ GeV.
The one on the right shows the technicolor signals in 
$e^+e^-\to \pi_T\pi_T$ and $e^+e^-\to W_L\pi_T$ expected if 
$M_{\pi_T}$ = 99 GeV/c$^2$, $M_{\rho_T}$ = 220 GeV/c$^2$ and $N_D$ = 9,
normalised to the integrated luminosity collected at 
$\sqrt{s}=192-208$ GeV.}
\label{fig:mass}
\end{center}
\end{figure}

The distributions of some discriminating variables for data, 
the SM prediction, and technipion production are shown in Fig.~\ref{fig:pre4j}. 
The mass $M_{5C}$ of the jet pair after the 5C fit 
for the pairing with the smallest $\chi^2_{5C}$ 
is used as the $\pi_T$ mass estimator. 
Figure ~\ref{fig:pre4j-mass} shows its distribution for preselected events, 
for the Standard Model (SM) background sources, and for technipion production 
with $M_{\pi_T} = 99$ GeV/c$^2$. The possible contribution of $\pi_T \pi_T$ 
production would be seen as a narrow peak. The channel $W_L \pi_T$ would give 
a slightly wider peak shifted towards the mass of the W. The form of 
the mass spectrum of the sum of these two channels depends on the 
$\rho_t$ mass and the mixing angle $\chi$ (see Eq. (\ref{eq2})). This 
figure also shows the distribution of the final discriminant variable 
from the neural network output.
Figure \ref{fig:eff} shows the number of selected events as a function of the 
efficiency for a $\pi_T \pi_T$ signal, which is varied by changing
the cut on the NN output. The dependence is shown separately for the two years 
of data taking used. Figure \ref{fig:mass} shows the $M_{5C}$ mass spectrum 
for events with the NN output greater than 0.30 for the full 
statistics collected at $\sqrt{s}=192-208$ GeV. A reasonable agreement 
between data and the SM prediction is observed in all distributions,
the remaining differences are included in the systematic errors.

Figure \ref{fig:mass} also shows the expected spectrum 
of $W_L\pi_T$ and $\pi_T\pi_T$ production 
for $M_{\pi_T}$=99 GeV/c$^2$, $M_{\rho_T}$=220 GeV/c$^2$ and $N_D=9$
normalised to the collected luminosity. For these model parameters
the signal to background ratio for events with $M_{5C}>96$ GeV/c$^2$
is about 6.

In addition to the NN analysis, a 
sequential analysis was also developed. Its performance is slightly
worse, and therefore it is used only as a cross-check.
After the preselection stage it uses three discriminating variables. 
Two of them are intended to reduce the $q\bar{q}(\gamma)$ contamination. 
They are $y_{34}$, defined above, 
and the sum of the second and fourth Fox-Wolfram moments, $H_2+H_4$.
Events are required to have $y_{34}>0.003$ and $H_2+H_4<0.6$.
The cut on the ${\mathrm b}$-tagging variable $x_b>1.3$
is used to suppress the $W^+W^-$ background.

Tables~\ref{tabdb200} and \ref{tabdb207} give the numbers of selected and
expected events at different steps of the sequential analysis together
with the efficiency of the signal selection. For comparison, the
results of the NN analysis for NN output cuts giving
similar signal efficiencies are also shown. The results of both analyses
show good agreement of the data with the SM prediction. No contribution
from technicolor production is observed.

\begin{table}[htbp]
\centering
\begin{tabular}{|c|c|c|c|c|c|c|c|c|}
\hline
channel & \multicolumn{8}{c|}
    {$M_{\pi_T}$ (GeV/c$^2$)  } \\ \cline{2-9}
        &  50 & 60 & 70 & 80 & 90 & 99 & 100 & 110 \\
\hline
$W_L \pi_T$   &  7.9     &  9.5   & 11.0 &  11.5 & 12.9 & &  14.6 &  13.9                \\
$\pi_T \pi_T$ & 23.7   &  32.9 &  33.9 & 36.0 & 42.5 & 49.6 & &              \\
\hline
\end{tabular}
\caption[]{Search in the 4-jet final state: selection efficiency in percent 
(including topological branching ratios) for $W_L \pi_T$ and $\pi_T \pi_T$ 
for different $\pi_T$ masses $M_{\pi_T}$, \rs~=~200~\GeV, and NN output 
variable $>0.3$.}
\label{tabs200}
\end{table}

\begin{table}[htbp]
\begin{center}
\begin{tabular}{|c|c|c|c|c|c|c|}     \hline
  Selection     & Data & Total             & \qqg  & 4 fermion & Efficiency &
 Efficiency \\
   &      & background        &       &           & $\pi_T \pi_T$ (\%)  &
$W_L \pi_T$ (\%) \\ \hline
Preselection   & 2455 & 2471.4 &   751.7  &  1719.7 &    93.4  & 62.5 \\
$y_{34} \geq 0.003$   & 2035 & 2042.4& 460.3&  1582.1&    90.0 & 58.6 \\
$H_2+H_4 \leq 0.6 $    &  1459 &  1488.1&  178.2 &  1309.9 &  78.5 & 51.7\\
$x_b\geq$ 1.3   &   48 & 50.0 & 20.8  &  29.2 &    43.9 & 14.3 \\
 \hline
NN$>0.3$   &   32 & 37.6 & 12.4  &  25.2&    42.5 & 12.9 \\
\hline
\end{tabular}
\caption[] {Search in the 4-jet final state: 
effect of the selection cuts in the sequential analysis on data, 
simulated  background and simulated signal events at \rs~=~192-202~\GeV.
Efficiencies are given for $M_{\pi_T}~=~90$ GeV/c$^2$
and include the topological branching ratios of $W$ and $\pi_T$ to two jets.}
\label{tabdb200}
\end{center}
\end{table}

\begin{table}[htbp]
\begin{center}
\begin{tabular}{|c|c|c|c|c|c|c|}     \hline
  Selection     & Data & Total             & \qqg  & 4 fermion & Efficiency &
 Efficiency \\
   &      & background        &       &           & $\pi_T \pi_T$ (\%)  &
$W_L \pi_T$ (\%) \\ \hline
Preselection   & 2266 & 2342.1 &   680.3  &  1661.8 &    91.1  & 64.9 \\
$y_{34} \geq 0.003$   & 1929 & 1940.7& 416.8&  1523.8&    89.3 & 60.7 \\
$H_2+H_4 \leq 0.6 $    &  1368 &  1395.6&  163.0 &  1232.7 &  72.8 & 52.6\\
$x_b\geq$ 1.3   &   43 & 46.4 & 18.1  &  28.3&    44.9 & 13.7 \\
 \hline
NN$>$0.34  &   29 & 30.2 &  9.3  &  20.9&    45.0 & 11.0 \\
\hline
\end{tabular}
\caption[] {Search in the 4-jet final state: 
effect of the selection cuts in the sequential analysis on data, 
simulated  background and simulated signal events at \rs~=~204-208~\GeV.
Efficiencies are given for $M_{\pi_T}~=~99$ GeV/c$^2$
and include the topological branching ratios of $W$ and $\pi_T$ to two jets.}
\label{tabdb207}
\end{center}
\end{table}


\subsection[]{Search in the Semileptonic Final State}

The search for the technipion is also performed in channels containing 
two quarks, a lepton and a neutrino, corresponding to the decays 
$W_L^+ \pi_T^- \to l^+ \nu q \bar{q}$ and $\pi_T^+ \pi_T^- \to 
\tau \bar{\nu} q \bar{q}$.
This final state is selected in two steps. 

Since the topology searched for is very close to that of semileptonic 
$W^+ W^-$ decays, a similar selection~\cite{ww-cs}
is applied at the first step. However, variables strongly correlated 
with the boson mass are not used, making the analysis efficient for a wide 
range of $\pi^-_T$ masses. 

Firstly loose initial cuts, requiring at least 7 charged particles,
transverse energy greater than $0.25 \sqrt{s}$, less than 30 GeV in a 
$30^\circ$ cone around the beam, and the polar angle of the missing momentum 
fulfilling $|\cos{\theta_{miss}}|<0.985$, are used to remove a large fraction
of the leptonic, $q\bar q(\gamma)$ and $\gamma\gamma$ events.

Then an isolated lepton candidate has to be found. 
The isolation criterion is defined 
in terms of the product $p \cdot \theta_{iso}$, where 
$p$ is the lepton momentum and $\theta_{iso}$ is the isolation angle between 
the lepton and the nearest charged particle with momentum greater than 
1~GeV/c. 
Electrons and muons are identified
using the standard DELPHI tools~\cite{delphi}
and $p \cdot \theta_{iso}$ is required to be above 250 GeV/c$\cdot$degrees. 
Any other isolated electron or muon with energy between 5 and 25 GeV or an 
isolated charged hadron or low multiplicity jet (less than 5 charged 
particles) is identified as a $\tau$-lepton candidate. For these, 
since some part of the tau energy is taken away by neutrinos, 
the isolation requirement is relaxed to $p\cdot\theta_{iso}>150$ 
GeV/c$\cdot$degrees.

\begin{figure}[tbh]
  \begin{center}
    \psfig{figure=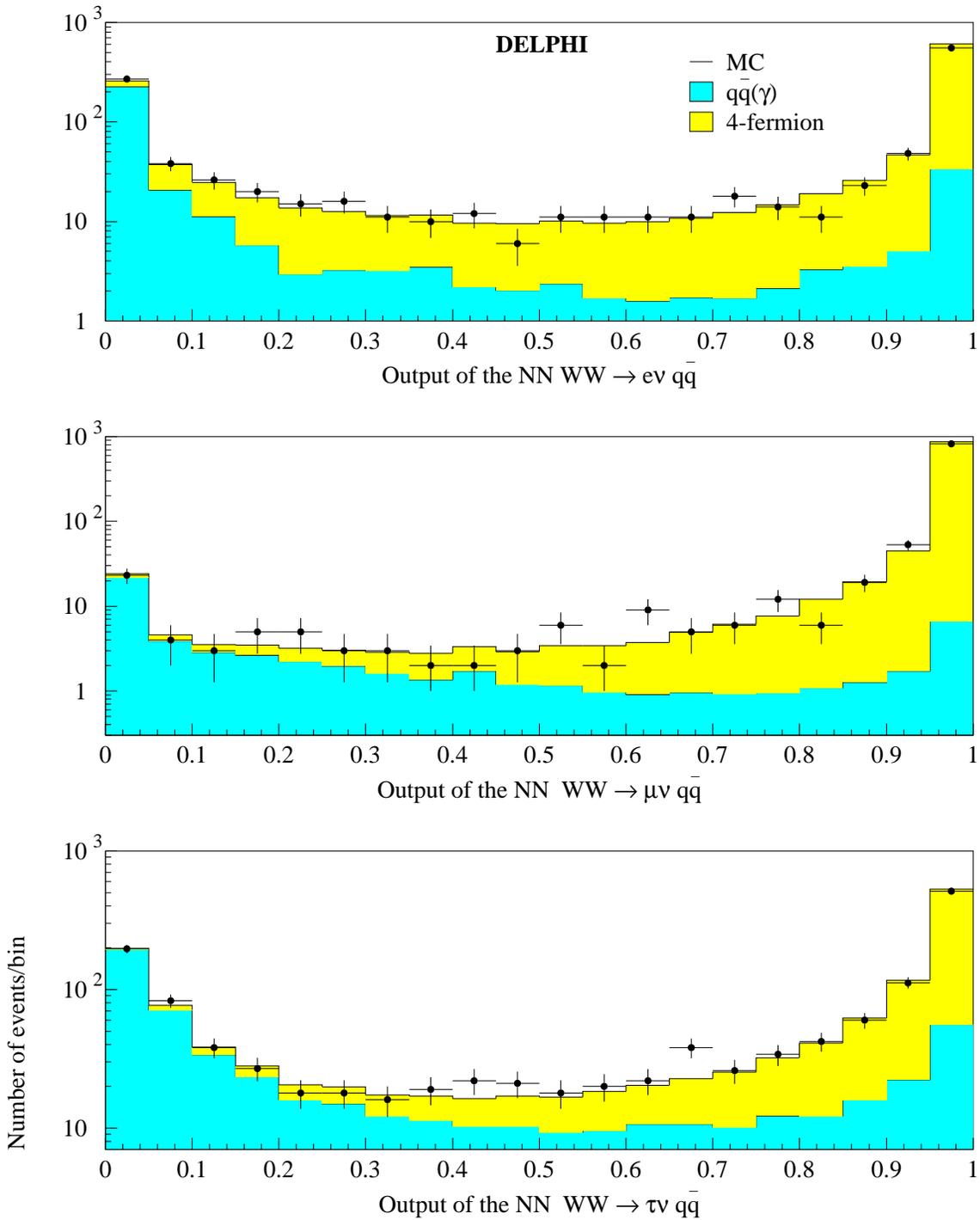,width=17cm}
    \caption[]{Search in the semileptonic final state:
    neural network outputs for the rejection of non-WW backgrounds
    for events with an electron candidate (top), a muon
    candidate (centre), or a tau candidate (bottom). 
      }
    \label{fig:nnww}
  \end{center}
\end{figure}

Depending on the flavour of the isolated lepton candidate, 
different neural networks are then used to reduce the background further. 
For a muon candidate, a neural 
network with 7 input variables is used: the lepton momentum, lepton isolation, 
missing momentum, $|\cos{\theta_{miss}}|$, transverse momentum, visible 
energy, and $\sqrt{s'/s}$ where $s'$ is the reconstructed
effective centre-of-mass energy~\cite{abreu}.
One more variable, the acoplanarity angle\footnote{
For any two vectors the acoplanarity is defined as the angle between
their projections on the plane perpendicular to the beam direction.}
between 
the lepton and the hadronic system, is used for an electron. 
For tau candidates, the missing momentum and visible
energy are less discriminant and are replaced by four new variables:
the thrust, the angle between 
the lepton and hadronic system, and the acoplanarity and acollinearity of the
hadronic jets. The neural network outputs for the different leptons are shown in 
Figure~\ref{fig:nnww}. The events are accepted if the NN value is above 
0.4 for electrons and muons and above 0.6 for taus. 
In this way most of the non-$W^+ W^-$ background is rejected. 

\begin{figure}[htbp]
  \begin{center}
    \psfig{figure=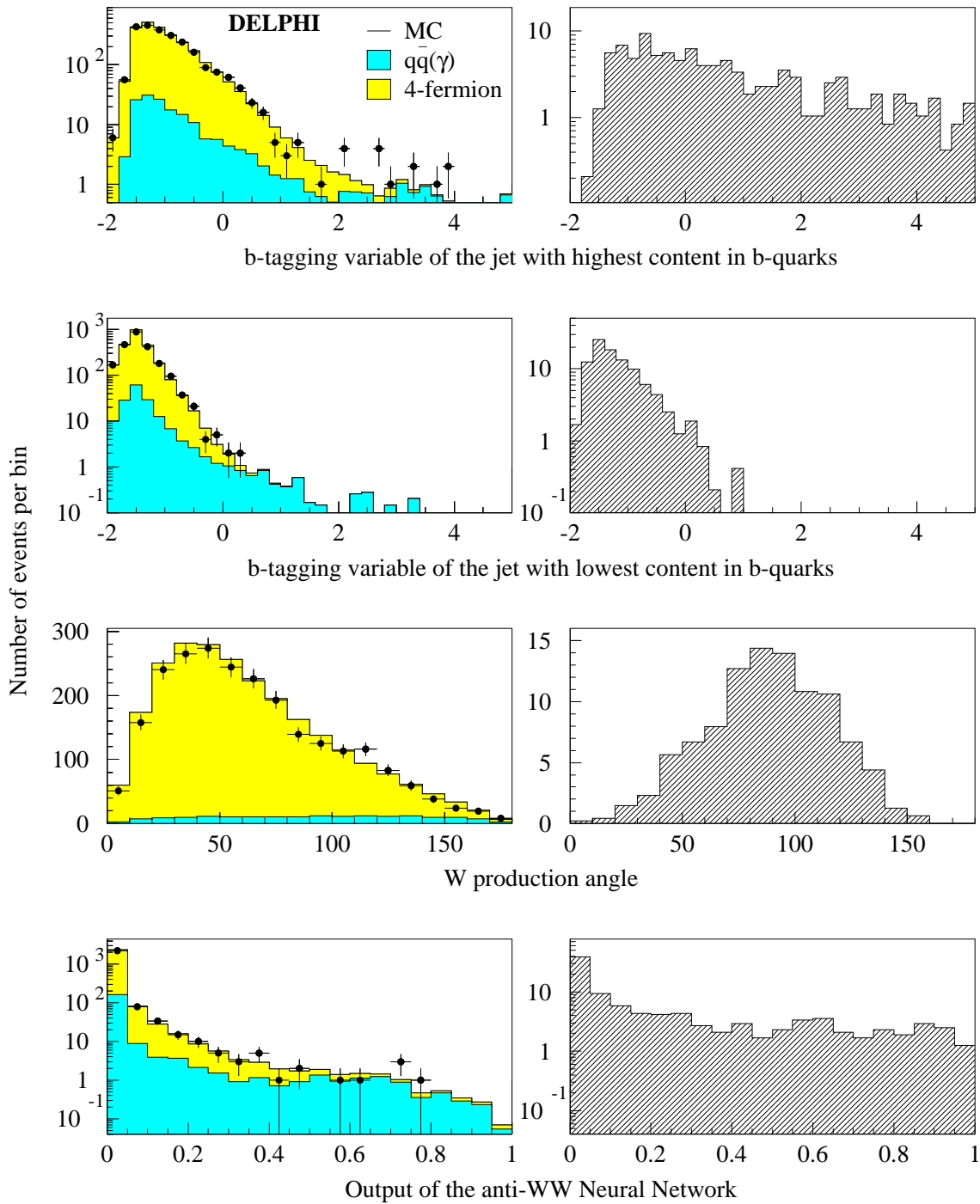,width=17.cm} 
    \caption[]{Search in the semileptonic final state:
      distributions after the rejection of non-WW
      background. The plots on the left show the data (points)
      and the expected SM backgrounds (histograms) for the full 
      DELPHI statistics at $\sqrt{s}=192-208$ GeV.
      Those on the right show the technicolor signal in 
      $e^+e^-\to W\pi_T$ expected if $M_{\pi_T}$ = 100 GeV.
      The signal normalisation corresponds to 
      $M_{\rho_T}=220$ GeV/c$^2$, $N_D=9$ and the integrated luminosity
      collected at $\sqrt{s}$=192-208 GeV.}
    \label{fig:varsbclv}
  \end{center}
\end{figure}

The second step exploits the specific properties of
the signal, such as the presence of b-quarks or the production angle, to
distinguish it from the W pairs.
This is done using another neural network which uses four input variables: 
the $b$-tagging variables of the two hadronic jets, 
$q \cdot \cos{\theta_{prod}}$ and $|\cos{\theta_{miss}}|$. The 
charge $q$ is defined according to that of the lepton, 
and the production polar angle $\theta_{prod}$ is built from the 
hadronic jets. The distribution of the $b$-tagging variable and
$q \cdot \cos{\theta_{prod}}$, together with the NN output are shown in 
Figure~\ref{fig:varsbclv}. 

This analysis provides good background suppression
and a reasonable selection efficiency of the $W_L \pi_T$ final state.
The $\pi_T \pi_T$ efficiency is limited
by the small $\pi_T \to \tau \bar{\nu}$ decay rate. Table~\ref{sl:tab1} gives
the $\pi_T \pi_T$ and $W_L \pi_T$ efficiencies for different $M_{\pi_T}$
masses obtained when selecting events with NN output $>0.1$.

\begin{figure}[tbh]
  \begin{center}
    \psfig{figure=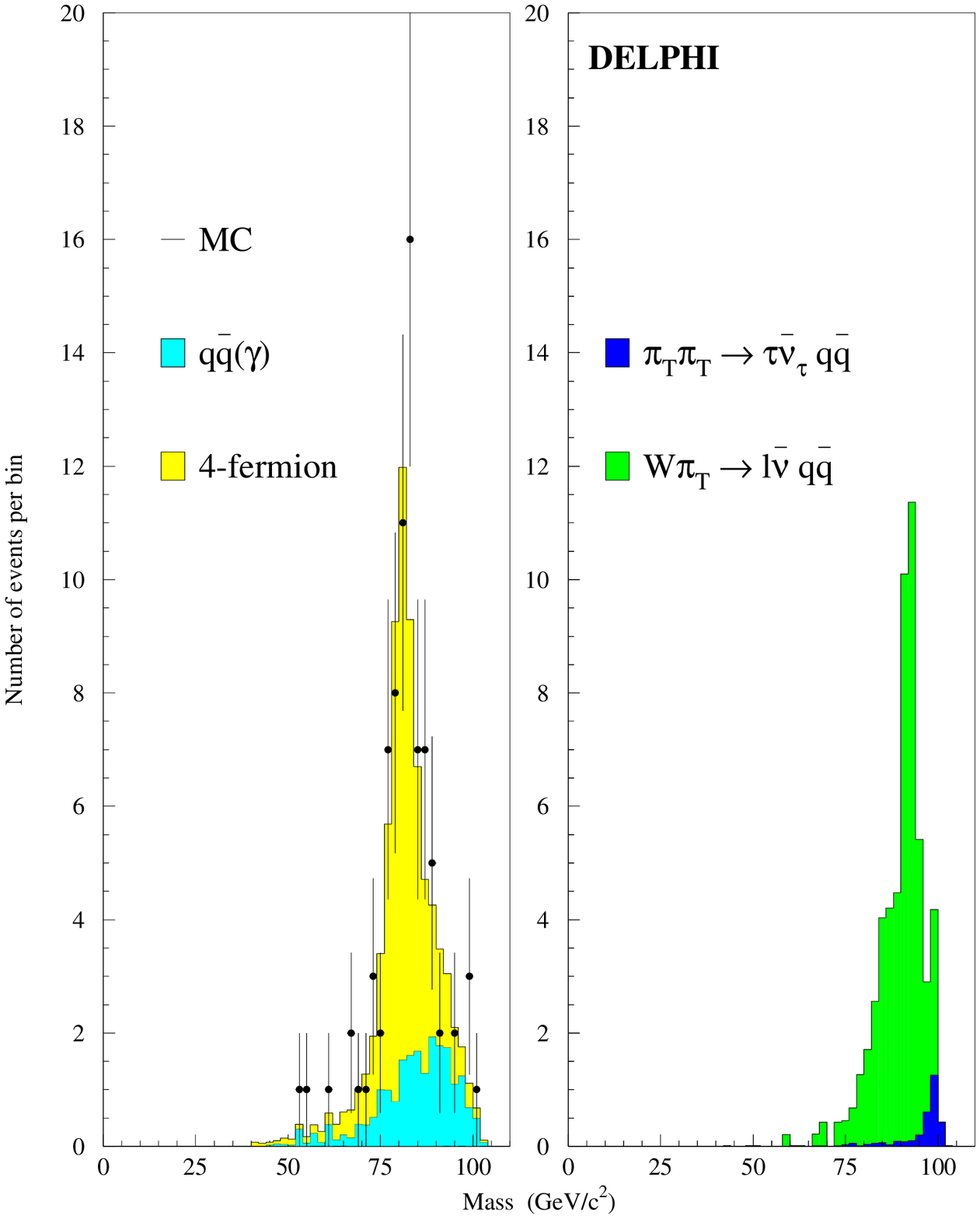,width=17cm}
    \caption[]{ Search in the semileptonic final state:
Estimated $\pi_T$ mass distributions for NN output $>0.10$. 
The plot on the left shows the data (points) and the expected SM backgrounds
(histograms) for the full DELPHI statistics at $\sqrt{s}=192-208$ GeV.
The one on the right shows the technicolor signals in 
$e^+e^-\to \pi_T\pi_T$ and $e^+e^-\to W_L\pi_T$ expected if 
$M_{\pi_T}$ = 100 GeV/c$^2$, $M_{\rho_T}$ = 220 GeV/c$^2$ and $N_D$ = 9,
normalised to the integrated luminosity collected at 
$\sqrt{s}=192-208$ GeV.}
    \label{fig:tclepmass}
  \end{center}
\end{figure}

The $M_{\pi_T}$ mass estimator is the same as in the hadronic channel.
The constrained fit is done with three additional free parameters 
coming from undetected neutrino for electron and muon, 
and with four parameters for tau, since also its energy is not known. 
Figure~\ref{fig:tclepmass} shows
the $\pi_T$ mass spectrum for events with the NN output greater
than 0.1 for the full statistics collected at $\sqrt{s}$=192-208 GeV.
This figure also shows the expected spectrum of $W_L \pi_T$
and $\pi_T \pi_T$ production for $M_{\pi_T}$=100 GeV/c$^2$,
$M_{\rho_T}$=220 GeV/c$^2$ and $N_D = 9$ normalised to the collected
luminosity. A good agreement between data and the SM prediction is
observed. 

Table~\ref{sl:tab2} gives the number
of selected and expected events at different steps of analysis
and for several cuts on the NN output. 
No contribution from technicolor production is observed. 

\begin{table}[htbp]
\centering
\begin{tabular}{|c|c|c|c|c|c|c|c|c|c|}
\hline
channel & \multicolumn{9}{c|}
    {$M_{\pi_T}$ (GeV/c$^2$)  } \\ \cline{2-10}
        &  50 & 60 & 70 & 80 & 90 & 99 & 100 & 110 & 120\\
\hline
$W_L \pi_T$   & 12.4   &  11.5 &  12.5 & 14.1 & 14.1 &      & 12.9 & 11.9 & 10.4   \\
$\pi_T \pi_T$ &  2.0   &   2.6 &   2.7 &  3.0 &  2.9 &  2.2 &      &      &        \\
\hline
\end{tabular}
\caption[] {Search in the semileptonic final state:
Selection efficiency in percent
(including topological branching ratios) for $\pi_T W_L$ and 
$\pi_T \pi_T$ for different $\pi_T$ masses $M_{\pi_T}$, 
$\sqrt{s}$ = 200 GeV, and NN output $>0.1$.}
\label{sl:tab1}
\end{table}

\begin{table}[htbp]
\begin{center}
\begin{tabular}{|c|c|c|c|c|c|}     \hline
  Selection     & Data & Total             & 
 $WW \rightarrow q \overline{q}' l \nu$ & \qqg & 
 Efficiency \\
   &      & background        &       &           & 
$W_L \pi_T$ (\%) \\ \hline
Hadronic preselection            &  19994 & 19626.1 & 2952.9 &    12446.3 &
96.9\% \\
$q\overline{q}' l \nu$ selection &   2375 &  2504.9 & 2309.1 &       63.1 &
23.5\% \\
NN output $> 0.1$                &     81 &    76.9 &   54.9 &        7.4 &
12.9\% \\
NN output $> 0.2$                &     32 &    33.2 &   18.8 &        5.3 &
10.4\% \\
NN output $> 0.3$                &     17 &    18.9 &    8.2 &        4.1 &
    7.4\% \\
\hline
\end{tabular}
\end{center}
\caption[]{Search in the semileptonic final state:
Effect of the selection cuts on data, 
simulated  background and simulated signal events 
at $\sqrt{s}$~=~192-208~GeV.
Efficiencies are given for $\pi_T W_L \to bc W_L $ with $M_{\pi_{T}}~=~100$ GeV/$c^2$.}
\label{sl:tab2}
\end{table}

\subsection{Combined result of the \boldmath $\pi_T$ search}

\label{sec4.3}

Since good agreement between data and the Standard Model
expectation is observed, the results are used to set limits 
on technicolor production, which are presented as 
a 95\% CL exclusion region 
in the $(M_{\rho_T}, M_{\pi_T})$ plane. 
The observed and expected limits quoted are based on the confidence level 
for signal, CL$_s$, as described in~\cite{read}. 
The test statistic used is a likelihood ratio, based on comparing the 
observed and expected rates and distributions 
as a function of mass and NN output.
The statistical and systematic errors on the expected background 
and signal distributions are taken into account. 

In the four-jet channel the relative systematic error was estimated 
at 11\% in the background level and 5\% in the signal efficiency. 
The main contribution, evaluated at about 10\% in 
the background and at 4\% in the signal efficiency,
comes from the $b$-tagging. In the semileptonic channel the main
uncertainty is related to the lepton identification efficiency.
The total relative error is estimated at 10\% in the background
and 2\% in the signal efficiency.

The $\pi_T \pi_T \to \tau \bar{\nu} q \bar{q}$ channel was not included
in the limits estimate, because its selection efficiency 
is significantly less than in the $\pi_T \pi_T \to q \bar{q} q \bar{q}$
channel, see tables \ref{tabs200},\ref{sl:tab1}.

\begin{figure}[tbh]
  \begin{center}
    \psfig{figure=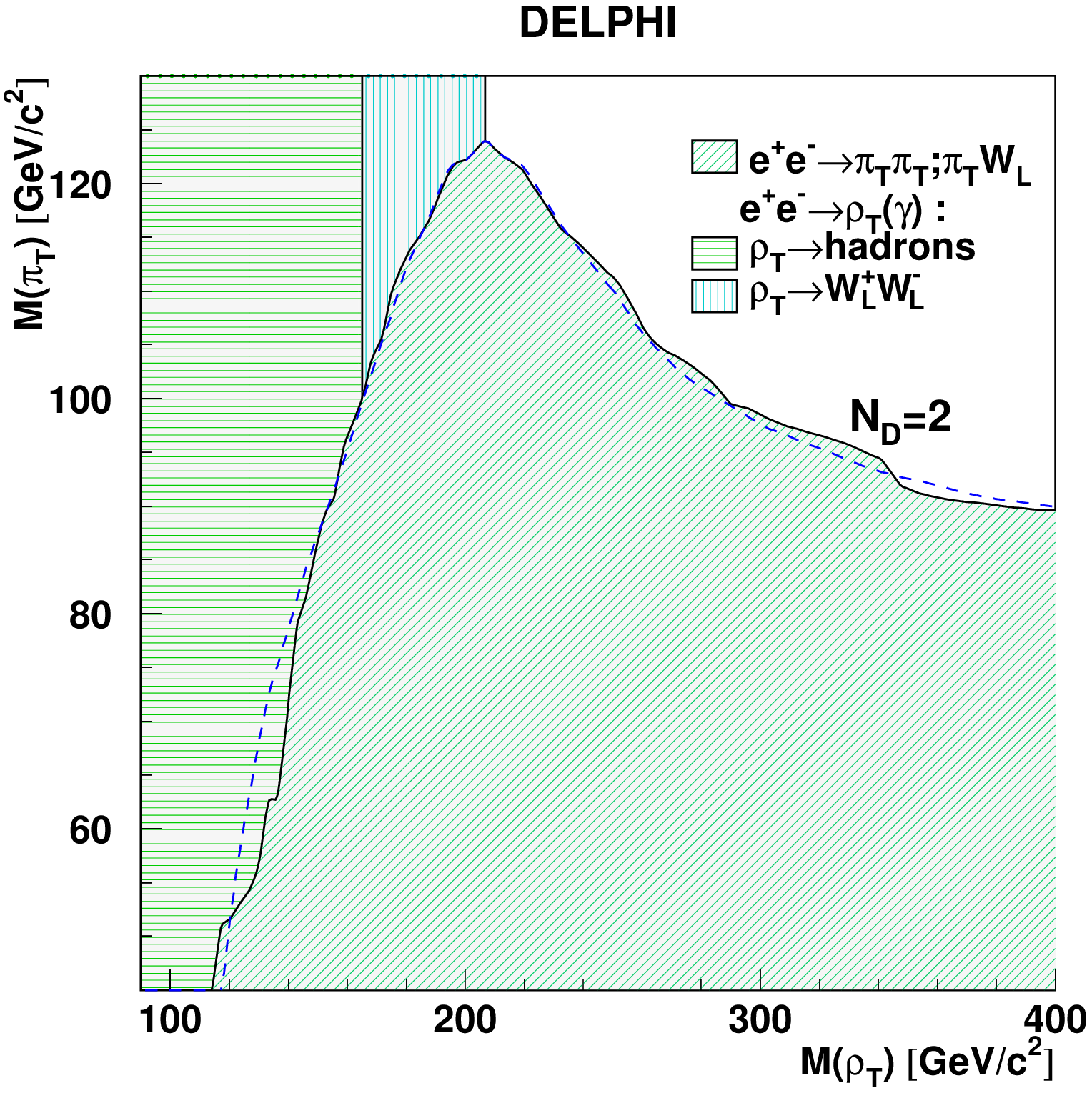,width=17cm}
    \caption[]{ The region in the $(M_{\rho_T} - M_{\pi_T})$ plane
      (filled area) excluded at 95\% CL for $N_D=2$ (maximal $W_L$ - $\pi_T$
      mixing). The dashed line shows the expected limit for the 
      $e^+e^- \to \pi_T \pi_T, \pi_T W_L$ search.
      }
    \label{nd2}
  \end{center}
\end{figure}

\begin{figure}[tbh]
  \begin{center}
    \psfig{figure=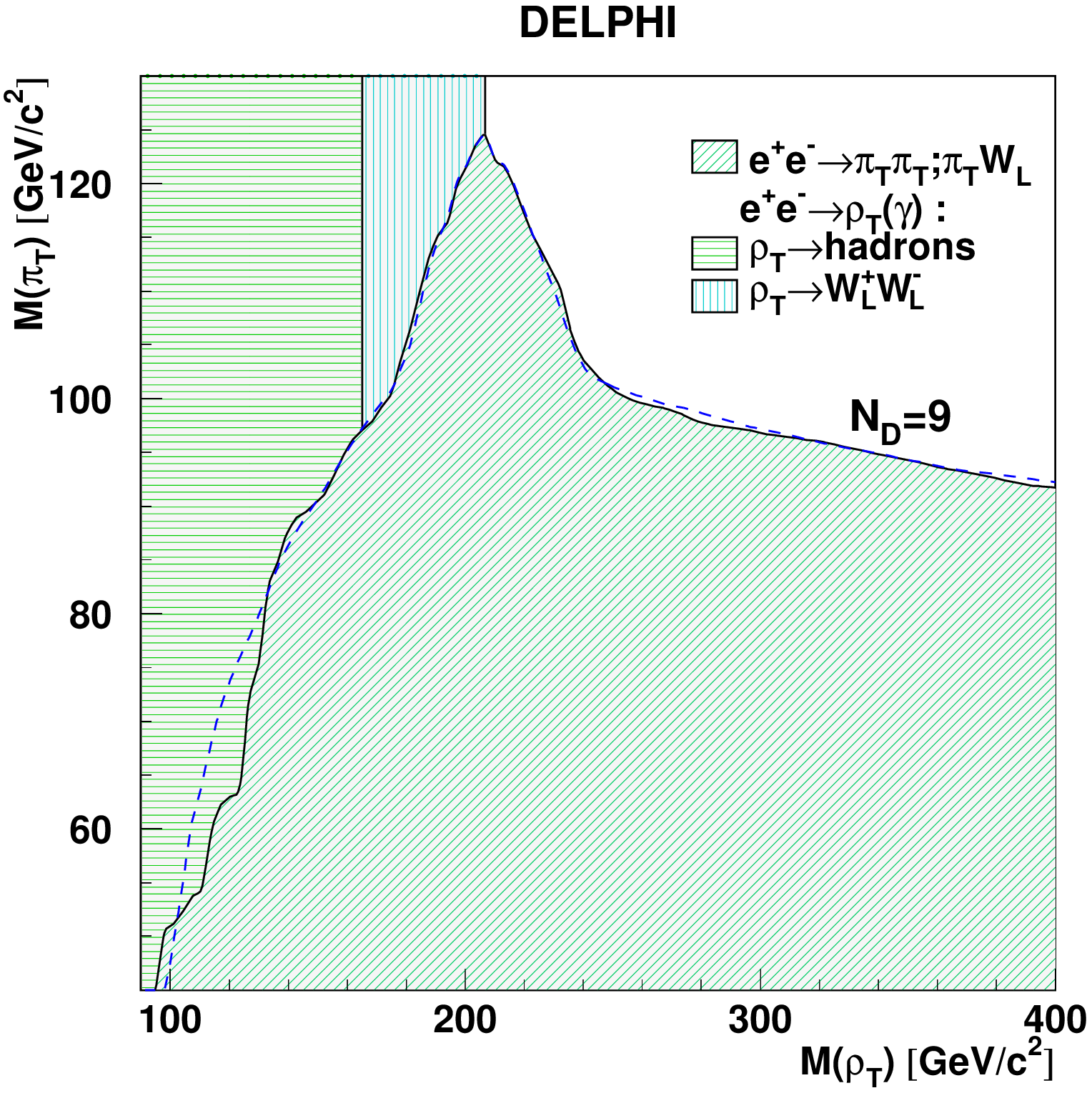,width=17cm}
    \caption[]{ The region in the $(M_{\rho_T} - M_{\pi_T})$ plane
      (filled area) excluded at 95\% CL for $N_D=9$ (theoretically
      preferred $W_L$ - $\pi_T$ mixing). The dashed line shows the 
      expected limit for the 
      $e^+e^- \to \pi_T \pi_T, \pi_T W_L$ search.
      }
    \label{nd9}
  \end{center}
\end{figure}

Two cases are considered separately, $N_D = 2$ (maximal mixing),
see Fig.~\ref{nd2}, and
$N_D = 9$ (theoretically preferred~\cite{lane}), see Fig.~\ref{nd9}.
The regions excluded by this analysis are shown by the diagonal hatching.

{\it In the limit of infinite} $\rho_T$ mass and assuming 
a point-like coupling of the gauge bosons to $\pi^+_T\pi^-_T$, the 
DELPHI data set 95\% CL lower limits on the charged technipion mass of 
$M_{\pi_T}$ = 79.8 GeV/c$^2$ (81.1 GeV/c$^2$ expected) for $N_D$ = 2, 
and $M_{\pi_T}$ = 89.1 GeV/c$^2$ (88.1 GeV/c$^2$ expected) for $N_D$ = 9.

Although the limit on the $\pi_T$ mass excludes a technicolor 
interpretation of the excess of events observed by L3~\cite{l3hh} 
at 68 GeV/c$^2$ in their $H^+H^-$ analysis, it should be noted that the 
DELPHI mass limit was obtained by applying $b$-tagging and therefore 
the present analysis cannot be compared directly with the L3 result.

Relaxing the hypothesis of a dominant technipion decay
into $b$ quarks, used in this analysis, will not drastically modify the 
obtained result for $N_D$=9. In this case the $\pi^+_T \pi^-_T$ channel,
when the $\rho_T$ becomes very heavy, has almost the same cross-section 
as the $H^+H^-$ channel of MSSM. Therefore, the results of $H^+H^-$ 
search~\cite{lephh}, which give a limit just below the W mass,
can also be used to set a limit on the technipion production. 
However, for $N_D$=2 the drop in production cross-section is significant
and a special analysis is required.

\section[]{Search for \boldmath $\rho_T$ with $M_{\rho_T} < \sqrt{s}$}

\label{sec2}

A $\rho_T$ with mass below $\sqrt{s}$ can be produced on mass
shell in the radiative return process $e^+e^- \to \rho_T (\gamma)$
with subsequent decay into different final states. 
This section presents the search for $\rho_T$ in all the main 
$\rho_T$ decay modes in the $M_{\pi_T}$ region not covered by the 
results of the section \ref{pipi}. It is based on a special search 
for the $\pi_T\gamma$ channel and on previous DELPHI 
measurements~\cite{ww-cs,qq-cs} of the $WW$ and $q\bar q$ 
production cross-sections.

\subsection[]{\boldmath $e^+e^- \to \rho_T (\gamma)$ 
with $\rho_T \to \pi_T^0 \gamma$}

\label{sec:pig}

The decay $\rho_T \to \pi^0_T \gamma$ is more favourable kinematically 
than charged $\pi_T$ pair production
and the dominant decay of $\pi^0_T$ into $b\bar{b}$ ($\sim$90 \%) allows 
a clean experimental signature. There is also an isosinglet 
called $\pi'^{0}_T$ which can decay into gluons and fermions and is 
expected to have about the same mass. To be conservative, its possible 
contribution is ignored. 

The hadronic events are selected by requiring at least 6 charged particles
with a total energy exceeding 24\% of the centre-of-mass energy. 
Any photon with an energy exceeding 5 GeV is considered as a possible
isolated photon candidate. All the other particles in the event 
are clustered into jets using the JADE algorithm~\cite{pyth},
and the photon is accepted as isolated if either its transverse momentum
to the nearest jet exceeds 10 GeV/c or the angle between its direction
and the nearest jet exceeds 45 degrees. More than one isolated photon
is allowed in an event.

A constrained fit requiring the conservation of energy and momentum
and allowing one additional photon in the beam pipe
is then applied to all selected events. An event is rejected if
the $\chi^2$ of this fit exceeds 9. The sum of all particles 
excluding the isolated photons is called the hadronic system.
The momentum of the hadronic system computed after 
the constrained fit is required to exceed 10 GeV/c, and the polar angle of 
its direction $\Theta_{had}$ to satisfy the condition 
$|\cos\Theta_{had}| < 0.9$. The reconstructed hadronic system is 
combined with the isolated photon, which is required to have  
$|\cos \Theta_{\gamma}| < 0.98$ where $\Theta_{\gamma}$ is
the polar angle of its direction. The energy of the combined 
(hadronic+photon) system is required to be less than $\sqrt{s}-5$ GeV, 
assuming at least one additional photon with energy above 5 GeV. 
Finally, as the main $\pi_T^0$ decay mode should be
$\pi_T^0 \to  b \bar{b}$, the b-tagging variable 
for the event $x_b$, defined in section \ref{sec41}, is required 
to exceed $-1$. The QCD background remaining after this cut has 
a $b$-purity of about 77\%.

\begin{figure}[tbh]
  \begin{center}
    \psfig{figure=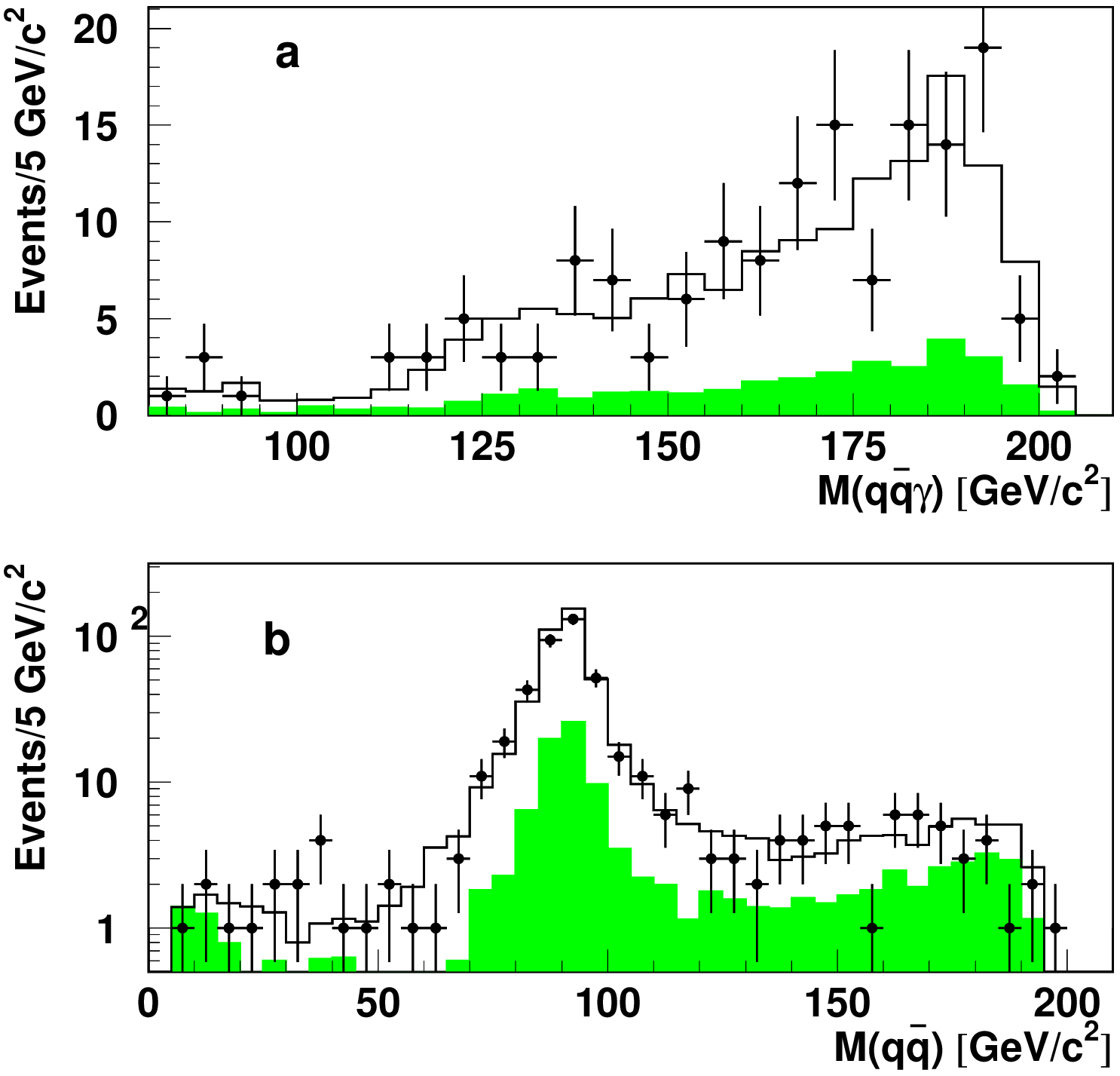,width=17cm}
    \caption[]{ $\pi_T^0 \gamma$ analysis: {\bf a)}
      distribution of the mass of the hadronic system plus
      the isolated photon; {\bf b)} distribution of the hadronic mass.
      The points show the data, the histogram shows the contribution 
      of standard sources, and the filled histogram shows separately 
      the contribution of all non-$b \bar{b} \gamma$ processes. The
      statistics shown in figures {\bf a)} and {\bf b)} corresponds
      to different event selections, see the text for details.
      }
    \label{fig:qqg}
  \end{center}
\end{figure}

With these selections 156 events are observed in the statistics collected
in 1999 and 2000 while 149.9 events are expected from the different SM sources.
Figure~\ref{fig:qqg}a shows the $(q \bar{q} \gamma)$ mass distribution of 
all selected events. The production of $\rho_T$ should manifest itself 
as a peak both in the distribution of the hadronic mass, corresponding
to the $\pi_T^0$, and in the mass of the hadronic system plus photon,
corresponding to the $\rho_T$, while no contribution from 
$\rho_T \to \pi_T^0 \gamma$ is seen in Fig.~\ref{fig:qqg}a.
A 15\% systematic error is assigned, which
takes into account the uncertainty in the selection efficiency 
of $b\bar{b} \gamma (\gamma)$ events (10\%) and uncertainty in the standard 
model cross-section $e^+e^- \to q \bar{q} \gamma (\gamma)$ (11\%).
Within the framework of the model~\cite{lane}, the resulting 
95\% CL upper limit on the branching ratio
$BR(\rho_T \to \pi_T^0 \gamma)$ does not exceed 7\% for 
$90 < M_{\rho_T} < 202 $ GeV/c$^2$. 

Due to this upper limit on $BR(\rho_T \to \pi_T^0 \gamma)$, the other
decay modes ($\rho_T \to W_L W_L$, $q \bar{q}$, $\pi_T \pi_T$) must dominate. 
The search for these channels is presented in the following sections.

In addition, the $\pi_T \gamma$ system can be produced in process
(\ref{eq3}), even if $M_{\rho_T} > \sqrt{s}$. The topology of this
process is different, and therefore the condition that the energy
of the (hadronic+photon) system is at least 5 GeV below $\sqrt{s}$ is
not applied. Dropping this condition, 468 events are selected in data 
and 502.6 events are expected from the standard sources. 
The distribution of the hadronic mass for this selection is shown in 
Fig.~\ref{fig:qqg}b, where only the expected $Z^0$ peak from the 
radiative return process is observed. 

The exclusion region in the ($M_{\rho_T}, M_{\pi_T}$) plane coming 
from the search for $e^+e^- \to (\rho_T^*,\omega_T^*) \to \pi_T^0 \gamma$ 
production is strongly model dependent and can even completely disappear
for $Q_U+Q_D = 0$ (see eq.~\ref{eq3}). In addition, for the typical 
parameter values, the extension of the limit given by other channels
is rather small. Therefore, the results of the 
$e^+e^- \to (\rho_T^*,\omega_T^*) \to \pi_T^0 \gamma$ search are not
included in the exclusion region given in Figs.~\ref{nd2},\ref{nd9}.



\subsection[]{\boldmath $e^+e^- \to \rho_T (\gamma)$ with $\rho_T \to W_L W_L$}

\label{sec:ww}

This section presents the search for the $\rho_T \to W_L W_L$
decay with the $\rho_T$ mass above the $2 M_W$ threshold.
It supposes that the $M_{\pi_T}$ value is not excluded by the analysis 
of section \ref{pipi} (see Figs.~\ref{nd2}, \ref{nd9}), 
i.e. that the channels $\rho_T \to W_L \pi_T$, $\pi_T \pi_T$ are 
kinematically closed. 

\begin{figure}[tbh]
  \begin{center}
    \psfig{figure=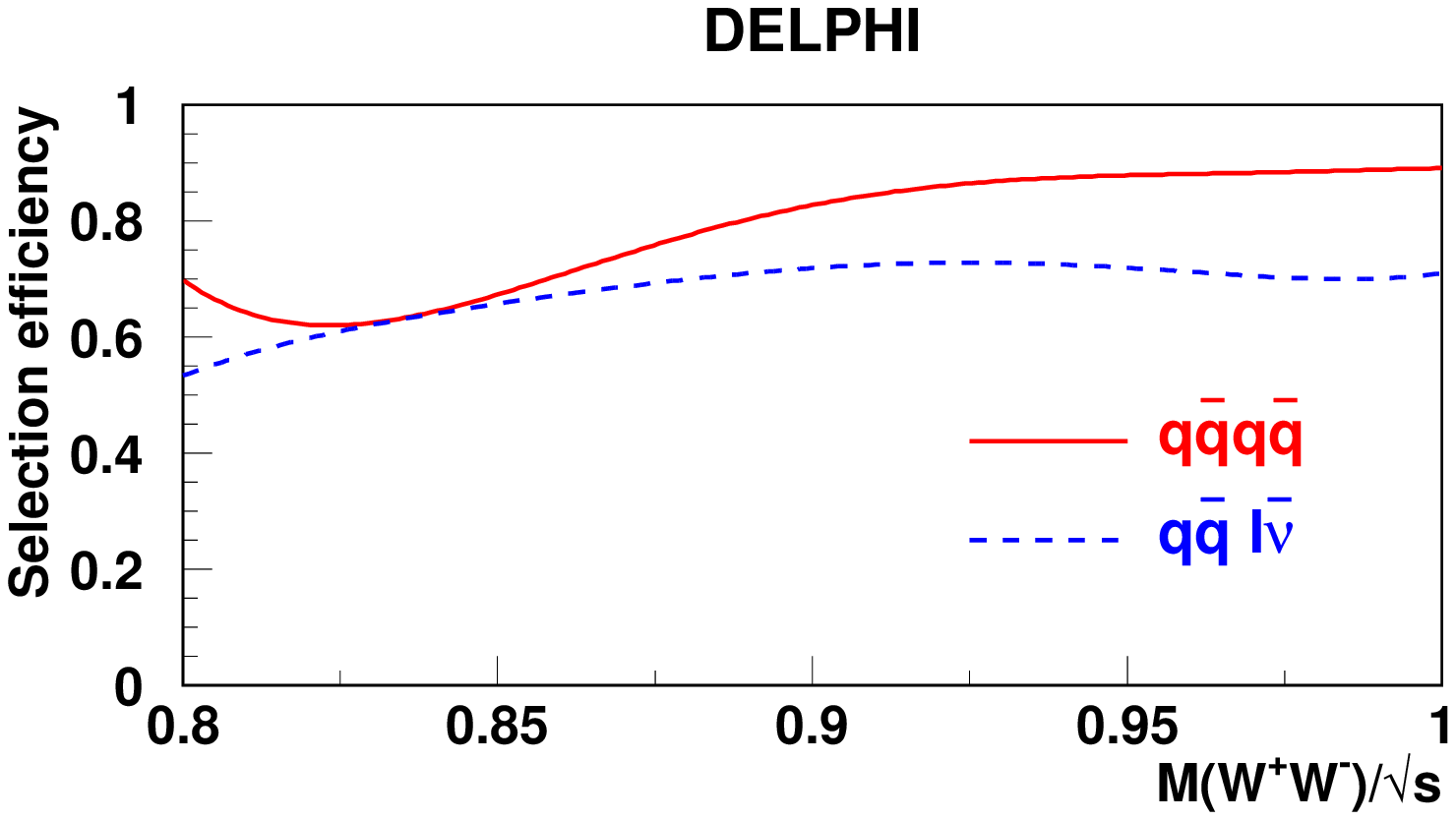,width=17cm}
    \caption[]{ Selection efficiency of a $WW$-like final state as a
      function of $M(W^+W^-)/\sqrt{s}$ for $\sqrt{s} = 206$ GeV.}
    \label{fig:tcww}
  \end{center}
\end{figure}

The search for this decay uses the DELPHI measurement of 
the $W^+ W^-$ cross-section at $\sqrt{s} = 172-206.7$ GeV~\cite{ww-cs},
which applies no strong condition on the energy of any ISR photon.
Figure \ref{fig:tcww} shows the resulting stability 
of the selection efficiency over wide ranges of $M_{W^+W^-}/\sqrt{s}$
for both the $q\bar{q}q\bar{q}$ and $q\bar{q}l\bar{\nu}$ final states.
Therefore the decay mode $\rho_T \to W_L W_L$ would give an additional 
contribution to the $W^+W^-$ cross-section. 

The measured values of the $W^+W^-$ cross-section are taken from~\cite{ww-cs}.
The Standard Model prediction is computed using the RacoonWW 
generator~\cite{Racoon}, while the selection efficiency is computed 
using EXCALIBUR~\cite{exc}. An additional 2\% systematic uncertainty 
is assigned to take into account a possible impact on the selection 
efficiency of differences in the event topology between these two 
generators. This analysis conservatively supposes all systematic errors 
to be fully correlated.
The expected cross-section of $e^+e^-\to \rho_T (\gamma)$ 
for some specific $\rho_T$ mass values is given in Table~\ref{tab1}.
The precision of $W^+W^-$ cross-section measurement is significantly better,
e.g. DELPHI reported $\sigma = 15.83 \pm 0.38 \pm 0.20$ pb at 
$\sqrt{s} = 189 $~GeV and the expected Standard Model
value is 16.25 pb.
 
No additional statistically significant contribution 
to the $W^+W^-$ cross-section is observed for any centre-of-mass energy.
Instead, the available measurements of the $W^+W^-$ cross-section 
put a 95\% CL upper limit on the branching ratio $BR(\rho_T \to W^+W^-)$.
It depends on the $\rho_T$ mass but in all cases is below 30\%.
Since $BR(\rho_T \to \pi_T^0 \gamma)$ is limited to 7\% at 95\% CL
(see section \ref{sec:pig}), the decay $\rho_T \to W_L W_L$ must 
be dominant in the $(M_{\rho_T},M_{\pi_T})$ mass region considered.
Therefore, the result obtained excludes
$\rho_T$ production for all $M_{\rho_T}$ between $2M_W$ and 206.7 GeV/c$^2$
and for all $M_{\pi_T}$ not excluded by the analysis of section \ref{pipi}. 
The region in the ($M_{\rho_T},M_{\pi_T}$) plane excluded by this analysis 
is shown by the vertical hatching in Figs.~\ref{nd2},\ref{nd9}.

\begin{table}[t]
  \centering
  \begin{tabular}{|c|l|l|l|l|l|l|l|l|}
    \hline 
    $M_{\rho_T}$ & \multicolumn{8}{c|} {$\sqrt{s} (GeV)$} \\
    \cline{2-9}
    (GeV/c$^2$) & ~183 & ~189 & ~192 & ~196 & ~200 & ~202 & ~205 & ~207 \\
    \hline
    175 & ~7.00~ & ~4.39~ & ~3.69~ & ~3.03~ & ~2.57~ & ~2.38~ & ~2.15~ & ~2.01~ \\
    185 & ~~--   & 10.68~ & ~7.25~ & ~5.06~ & ~3.87~ & ~3.45~ & ~2.97~ & ~2.71~ \\
    195 & ~~--   & ~~--   & ~~--   & 18.82~ & ~8.69~ & ~6.83~ & ~5.15~ & ~4.42~ \\ 
    \hline
  \end{tabular}
  \caption []{ Expected $e^+e^-\to\rho_T (\gamma)$ cross-section (in pb) 
    at different centre-of-mass energies for some $\rho_T$ mass values.}
\label{tab1}
\end{table}

\subsection[]{\boldmath $e^+e^- \to \rho_T (\gamma)$ with $\rho_T \to$ hadrons
$(q\bar{q},$ $\pi_T \pi_T)$}

For $M_{\rho_T} < \sqrt{s}$, technicolor production by process
(\ref{eq1}) would give a significant contribution to the cross-section
for $q\bar{q} (\gamma)$ production 
because the main $\rho_T$ decay channels all
include hadronic final states. 
Due to the relatively small $\rho_T$ decay width,
this contribution would be observed as a peak in the hadronic
mass distribution. The search for this decay channel uses all 
published DELPHI $q \bar{q} (\gamma)$ cross-section measurements,
which are currently available for $\sqrt{s}= 183$ and 189 GeV 
~\cite{qq-cs}, and is limited to $\rho_T$ mass values 
below 165 GeV/c$^2$. Above 165 GeV/c$^2$ either the decay 
$\rho_T \to W_L W_L$, considered in section \ref{sec:ww}, or the 
decays $\rho_T \to (\pi_T \pi_T$, $W_L \pi_T)$, considered in 
section \ref{pipi}, become dominant.

The topology of $\rho_T \to q \bar{q}$ events is almost the same as that
of standard $e^+e^- \to q \bar{q} (\gamma)$ processes, while the
decay $\rho_T \to \pi_T \pi_T$ produces many-jet events. However,
the $q \bar{q} (\gamma)$ selection criteria~\cite{qq-cs} 
are quite loose, allowing effective selection of both $\rho_T$ decay modes.
This was verified by passing simulated 
$e^+e^-\to\rho_T(\gamma)\to\pi_T\pi_T(\gamma)$ events through the complete 
$q\bar{q}(\gamma)$ analysis chain. The selection efficiency was found to be 
the same as for standard $q\bar{q} (\gamma)$ events.

\begin{figure}[tbh]
  \begin{center}
    \psfig{figure=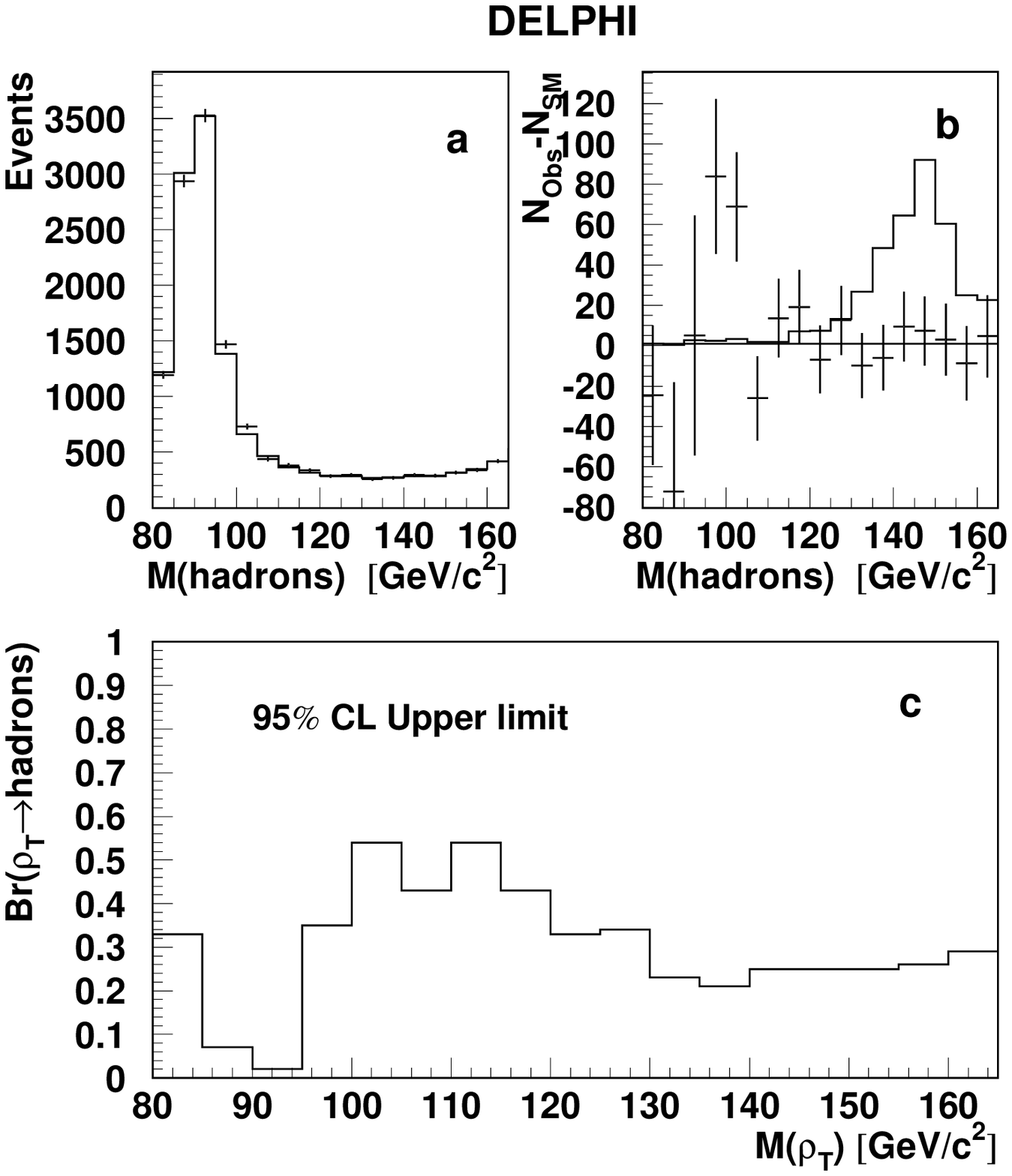,width=17cm}
    \caption[]{ {\bf a)} Mass distribution of the hadronic system in the
      $e^+e^- \to q \bar{q} (\gamma)$ analysis for the data collected
      at $\sqrt{s} = 183$ and 189 GeV. Crosses show the data and the 
      histogram shows the SM contribution. {\bf b)} Difference between
      the observed numbers of events and those expected in the SM.
      The expected contribution of $\rho_T \to \pi_T \pi_T$ with 
      $M_{\rho_T} = 150$ GeV/c$^2$ and $M_{\pi_T} = 70$ GeV/c$^2$ is shown 
      as the histogram. 
      {\bf c)} The 95\% CL upper limit on the branching ratio
      BR($\rho_T \to$hadrons).}
    \label{fig:tcqq}
  \end{center}
\end{figure}

Figure \ref{fig:tcqq}a shows the observed mass distribution of the 
hadronic system together with the expected 
contribution from Standard Model processes. The hadronic mass
reconstruction is described in~\cite{qq-cs}. Figure \ref{fig:tcqq}b
shows the difference between the observed and expected numbers of events
and the contribution of a $\rho_T \to \pi_T \pi_T$ signal with 
$M_{\rho_T} = 150$ GeV/c$^2$ and $M_{\pi_T} = 70$ GeV/c$^2$. Good sensitivity 
to technicolor production can be seen. 

Using the observed and expected numbers of events gives 
the 95\% CL upper limit on the decay branching ratio BR($\rho_T \to$hadrons)
shown in Fig.~\ref{fig:tcqq}c. The small mismatch between data
and simulation for the width of the radiative return to the $Z^0$ 
in Fig.~\ref{fig:tcqq}a is due to imprecise modeling of such details 
as jet angles and momenta. It explains some increase of the 
BR($\rho_T \to$hadrons) limit around 100 GeV, which, however, remains 
below 55\%. Taking into account that $BR(\rho_T \to \pi_T^0 \gamma)$ is 
limited by 7\% at 95\% CL (see sec. \ref{sec:pig}), this result excludes
$\rho_T$ production for all $\rho_T$ masses between 90 and 165 GeV/c$^2$.
The horizontal hatching in Figs.~\ref{nd2}, \ref{nd9} show the contribution 
of this channel in the combined excluded region in the 
$(M_{\rho_T},M_{\pi_T})$ plane. 



%

\section[]{Summary}
\label{results}

This paper presented the search for $\pi_T \pi_T$ and $W_L \pi_T$ production 
in process (\ref{eq2})
and for $\rho_T$ production in the radiative return process (\ref{eq1})
followed by the decays $\rho_T \to \pi^0_T \gamma$,
$\rho_T \to W^+W^-$ or $\rho_T \to$hadrons.
A good agreement between data and the Standard
Model expectation is observed in all channels studied. 
The combined region in the $(M_{\rho_T},M_{\pi_T})$ plane excluded 
by this analysis at a 95\% CL is shown in Figs.~\ref{nd2},\ref{nd9}.
A 95\% CL lower mass limit 
of 79.8 GeV/c$^2$ is set independently of other
parameters of the technicolor model, supposing 
its point-like coupling with gauge bosons (see section~\ref{sec4.3}).
The $\rho_T$ production is excluded at 95\% CL
for $90 < M_{\rho_T} < 206.7$ GeV/c$^2$ independently of the $\pi_T$
mass and all other model parameters.

These results significantly improve on the exclusion limits on technicolor
production obtained by the CDF experiment \cite{ref4}.

\subsection*{Acknowledgements}
\vskip 3 mm
We wish to thank K.Lane for answering many questions on technicolor
models and encouraging this work.

 We are greatly indebted to our technical 
collaborators, to the members of the CERN-SL Division for the excellent 
performance of the LEP collider, and to the funding agencies for their
support in building and operating the DELPHI detector.\\
We acknowledge in particular the support of \\
Austrian Federal Ministry of Education, Science and Culture,
GZ 616.364/2-III/2a/98, \\
FNRS--FWO, Flanders Institute to encourage scientific and technological 
research in the industry (IWT), Belgium,  \\
FINEP, CNPq, CAPES, FUJB and FAPERJ, Brazil, \\
Czech Ministry of Industry and Trade, GA CR 202/99/1362,\\
Commission of the European Communities (DG XII), \\
Direction des Sciences de la Mati$\grave{\mbox{\rm e}}$re, CEA, France, \\
Bundesministerium f$\ddot{\mbox{\rm u}}$r Bildung, Wissenschaft, Forschung 
und Technologie, Germany,\\
General Secretariat for Research and Technology, Greece, \\
National Science Foundation (NWO) and Foundation for Research on Matter (FOM),
The Netherlands, \\
Norwegian Research Council,  \\
State Committee for Scientific Research, Poland, SPUB-M/CERN/PO3/DZ296/2000
and SPUB-M/CERN/PO3/DZ297/2000 \\
JNICT--Junta Nacional de Investiga\c{c}\~{a}o Cient\'{\i}fica 
e Tecnol$\acute{\mbox{\rm o}}$gica, Portugal, \\
Vedecka grantova agentura MS SR, Slovakia, Nr. 95/5195/134, \\
Ministry of Science and Technology of the Republic of Slovenia, \\
CICYT, Spain, AEN99-0950 and AEN99-0761,  \\
The Swedish Natural Science Research Council,      \\
Particle Physics and Astronomy Research Council, UK, \\
Department of Energy, USA, DE--FG02--94ER40817, \\

\clearpage


\end{document}